\documentclass[twocolumn,usenatbib,useAMS]{mnras}

\def\lsim{~\rlap{$<$}{\lower 1.0ex\hbox{$\sim$}}}
\def\bsim{~\rlap{$>$}{\lower 1.0ex\hbox{$\sim$}}}

\def\flux{\ {\rm erg}\,{\rm s}^{-1}\ {\rm cm}^{-2}}
\def\lum{\ {\rm erg}\ {\rm s}^{-1}}

\def\apjl{Astrophys.\ J.\ Lett.}
\def\mnras{Mon.\ Not.\ R.\ Astron.\ Soc.}

\def\aap{Astron.\ Astrophys.}
\def\apj{Astrophys.\ J.}

\def\prl{Phys.\ Rev.\ Lett.}
\def\prd{Phys.\ Rev.\ D}

\def\physrep{Phys. Rep.}

\def\jcap{{JCAP\ }}

\def\ln{{\rm ln}}

\def\mathbi#1{\textbf{\em #1}}

\def\vk{\mathbi{k}}

\def\vx{\mathbi{x}}

\def\fnl{f_\text{NL}}
\def\bphi{b_\phi}

\usepackage{graphicx}	
\usepackage{amsmath}	
\usepackage{amssymb}	
\usepackage{multicol}        
\usepackage{bm}		
\usepackage{pdflscape}	
\usepackage{hyperref}
\usepackage{xcolor}
\usepackage[normalem]{ulem}
\usepackage{transparent}

\usepackage[english]{babel}

\usepackage{hyperref} 
\newcommand{\Gal}{{\small GALACTICUS}\,\,}

\usepackage[T1]{fontenc}
\usepackage{ae,aecompl}

\usepackage{newtxtext,newtxmath}

\title{Non-Gaussian assembly bias from a semi-analytic galaxy formation model}
\author[M. Marinucci et al.]{M. Marinucci,$^{1}$\thanks{E-mail: \href{mailto:marinucci@campus.technion.ac.il}{marinucci@campus.technion.ac.il}}
V. Desjacques,$^{1,2}$
A. Benson,$^{3}$
\\
$^{1}$Physics department, Technion, 3200003 Haifa, Israel%
\\
$^{2}$Asher Space Research Institute, Technion, 3200003 Haifa, Israel%
\\
$^{3}$Carnegie Observatories, 813 Santa Barbara Street, Pasadena, CA 91101, USA}

\date{Accepted XXX. Received YYY; in original form ZZZ}

\pubyear{2023}

\begin{document}
\label{firstpage}
\pagerange{\pageref{firstpage}--\pageref{lastpage}}
\maketitle

\begin{abstract}
We use $z=1$ mock galaxy catalogues produced with the semi-analytic code \Gal to study the dependence of the non-Gaussian bias parameter $\bphi$ on the mass assembly history of the host halos. We generate large sets of merger trees and measure the non-Gaussian assembly bias $\Delta\bphi$ for galaxies selected by color magnitude and emission line luminosities. 
For galaxies selected by $g-r$ color, we find a large assembly bias 
consistent with the analysis of Barreira et al. (2020) based on hydro-dynamical simulations of galaxy formation. 
This effect arises from the fact that a larger value of the normalization amplitude $\sigma_8$ implies a faster mass assembly (at fixed halo mass) and, therefore, older and redder galaxies. 
On the contrary, for galaxies selected by their H$\alpha$ luminosity, we do not detect a significant assembly bias, 
at least at $z=1$ and in the halo mass range 
$3\times10^{10} < M < 10^{12}\ M_\odot$ considered here.
This is presumably due to the fact that emission line strengths are mainly sensitive to the instantaneous star formation rate, which appears to depend weakly on $\sigma_8$ at $z=1$.
This indicates that the non-Gaussian assembly bias should be less of a concern for future emission line galaxy surveys.

We also investigate, for the first time, the sensitivity of the non-Gaussian assembly bias to a change in the parameters of the galaxy formation model that control the AGN and stellar feedback as well as the star formation rate. 
When these parameters change within a factor of two from their fiducial value, they induce variations up to order unity in the measured $\Delta\bphi$, but the overall trends with color or luminosity remain the same.
However, since these results may be sensitive to the choice of galaxy formation model, it will be prudent to extend this analysis to other semi-analytic models in addition to halo mass and redshift.


\end{abstract}

\section{Introduction}

Understanding the origin of the fluctuations in the initial conditions of the density distribution remains one of the most prominent questions of modern cosmology. 
A potentially detectable amount of primordial non-Gaussianity (PNG) can be produced when (at least) one of the assumptions of single field, slow roll inflation is violated \citep{1990PhRvD..42.3936S,1992PhRvD..46.4232F,1994ApJ...430..447G,2003NuPhB.667..119A,2003JHEP...05..013M,2004JCAP...10..006C,2011JCAP...11..038C,2011JCAP...05..014T}. In particular, multifield models of inflation \cite{1997PhRvD..56..535L,2003PhRvD..67b3503L} can generate a primordial bispectrum (3-point correlation function) which peaks in the squeezed limit. A detection of this bispectrum shape parametrized by $\fnl$ would thus rule out single field slow roll inflation.
The current best constraints on PNG of the local type is $\fnl=-0.9\pm5.1$ at 68$\%$ C.L. and comes from the temperature anisotropies of the cosmic microwave background~\citep{PlanckPNG}. In the coming years galaxy surveys (such as Euclid~\cite{EUCLID:2011zbd,Amendola:2016saw}, SPHEREx~\citep{Dore:2014cca}) are expected to give competitive constraints thanks to their large volumes, the redshifts covered and the higher number of tracers observed.

The clustering of large scale structure tracers is sensitive to a local PNG through a scale-dependent term proportional to the product $\fnl\bphi$, where $\bphi$ is the {\it non-Gaussian bias} \citep{dalal/etal:2008}. Therefore, the tightest constraints on $\fnl$ will be achieved if good priors on $\bphi$ are available. The peak-background split argument implies a relation between $\bphi$, the galaxy number density $\bar n_g$, and the normalisation amplitude $\sigma_8$ \citep{slosar/etal:2008}. 
However, while this prediction has been thoroughly tested at the level of dark matter halos, it has been used only recently to measure $\bphi$ for mock galaxies extracted from detailed hydrodynamical simulations. Assuming universality of the halo mass function, a "universal" relation $\bphi = 2\delta_c (b_1 - 1)$ holds between $\bphi$ and $b_1$, the large scale galaxy bias, where $\delta_c$ is the critical overdensity. The current analyses using galaxy clustering data are performed assuming this $\bphi(b_1)$ relation which is expected to hold only for halos selected by virial mass: ref.~\cite{DAmicoPNG} found $\fnl = -30 \pm 29$ at 68$\%$ C.L while ref.~\cite{CabassPNG} found $\fnl = -33 \pm 28$ at 68$\%$ C.L assuming $\bphi = 2\delta_c (b_1 - 0.55)$. As emphasized by e.g. \cite{slosar/etal:2008, reid/etal:2010,barreira:2020,barreira:2022b}, constraints on $\fnl$ are sensitive to the prior imposed on $\bphi$ (wrong priors can lead to large systematic biases in the inferred values of $\bphi$). The latter can significantly deviate from the $\bphi(b_1)$ expectation if there is significant assembly bias. 

In the last decade, much attention has been devoted to the \textit{assembly bias}, which is the fact that the clustering of dark matter halos and galaxies is not solely determined by the mass of the virialized halo at the formation epoch, but also by additional factors such as the formation history and environment. Assembly bias was first introduced by \cite{sheth/tormen:2004,gao/springel/white:2005}, who noticed that older dark matter halos preferentially reside in overdense regions and, therefore, tend to be more clustered than younger halos with the same halo mass. This led to the conclusion that there exist additional factors which influence the clustering of halos beyond their mass. Since then, numerous studies have attempted to quantify and understand the nature of assembly bias for dark matter halos and galaxies \citep[e.g.][]{gao/white:2007,wechsler/etal:2006,giocoli/etal:2007,jing/suto/mo:2007,keselman/nusser:2007,croton/etal:2007,desjacques:2008,codis/etal:2012,aung/cohn:2016,chaves/montero/etal:2016,miyatake/etal:2016,paranjape/padmanabhan:2017,paranjape/etal:2018,zehavi/etal:2018,shi/sheth:2018,ramakrishnan/etal:2019,hellwing/etal:2020}.


The first estimates of the assembly bias dependence of $\bphi$ \citep{slosar/etal:2008,reid/etal:2010} relied on the excursion set approach to halo formation~\cite[e.g.][]{bond/etal:1991,lacey/cole:1993} and on N-body simulations. In particular, \cite{reid/etal:2010} found that, at fixed halo mass, the non-Gaussian bias $\bphi$ depends strongly on the halo formation time, with older halos having a much larger $\bphi$ than younger ones. Recently, \cite{lazeyras/etal:2023} showed that $\bphi$ is also correlated with halo properties such as the spin or concentration. Furthermore, \cite{barreira/TNG:2020,barreira:2022a} measured $\bphi$ 
for mock galaxies extracted from the {\small Illustris TNG} suite of hydrodynamical simulations \citep{barreira/TNG:2020,Barreira:2021dpt,barreira:2022a} and found that their measurements can largely deviate from the $\bphi(b_1)$ relation, not only due to the convolution with the halo occupation distribution (HOD) but also because of assembly bias. While this may be a concern for the accuracy of $\fnl$-constraints, one could also take advantage of a large non-Gaussian assembly bias to construct galaxy samples which enhance the information on $\fnl$ \citep[see][for a recent study]{barreira&krause}.

In this paper, we use a semi-analytical approach to galaxy formation in order to study the impact of mass assembly on the non-Gaussian bias of $z=1$ galaxies. We consider two different galaxy properties: colour magnitudes and emission line strengths.
These two observables depend on the formation history of the galaxy in a different way. They are basic galaxy properties measured in current and future galaxy surveys such as {\small SDSS-IV/eBOSS}, {\small DESI}, {\small Euclid} or {\small SPHEREx} \cite[see, e.g.,][]{Dawson:2015wdb,EUCLID:2011zbd,DESI:2016fyo,DESI:2016igz,Dore:2014cca}. Although semi-analytical models \citep[or SAMs, see][for instance]{kauffmann/etal:1993,galform,croton/etal:2006,somerville/etal:2008,galacticus} of galaxy formation are not as realistic as full hydro-dynamical simulations, they are realistic and versatile enough to allow for a thorough study of the assembly bias as will be demonstrated below. Therefore, our study nicely complements the work of \cite{barreira/TNG:2020,barreira:2022b}, who relied on detailed hydro-dynamical simulations, and also extends it to include emission lines.
Our paper is organized as follows. We provide a quick theoretical overview and describe our methodology in \S\ref{sec:background}. We present our results in \S\ref{sec:results}, and conclude in \S\ref{sec:conclusions}.

\section{Theoretical background and methodology}
\label{sec:background}

\subsection{Primordial non-Gaussianity and galaxy bias}

Local primordial non-Gaussianity can be conveniently parametrized through the mapping  \citep{1990PhRvD..42.3936S,1994ApJ...430..447G,2000MNRAS.313..141V,2001PhRvD..63f3002K}
\begin{equation}
\label{eq:png}
\phi(\vx) = \phi_G(\vx) + \fnl \left[\phi_G(\vx)^2 - \left\langle \phi_G(\vx)^2 \right\rangle\right], 
\end{equation}
where the primordial gravitational potential $\phi(\vx) = (3/5) \mathcal{R}(\vx)$ is defined as Bardeen's curvature perturbation immediately after equality, $\phi_G$ is a Gaussian distributed random field, and $\fnl$ is a parameter that quantifies the level of non-Gaussianity of the spatial distribution of the primordial potential \citep{2001PhRvD..63f3002K}. The power spectrum of $\phi_G$ is $P_\phi(k)\propto k^{n_s-4}$, where $n_s<1$ is the scalar spectral index.

Large scale structure (LSS) tracers are especially sensitive to local primordial non-Gaussianity and can thus be used to constrain $\fnl$ \citep[see][for a recent review]{biagetti:2019}. In particular, local PNG introduces a coupling between large and small scales in the primordial fluctuations which leaves a broadband, scale-dependent imprint in the Fourier modes of LSS tracers. This leads to a scale-dependent bias at linear level \citep{dalal/etal:2008,slosar/etal:2008,matarrese/verde:2008,mcdonald:2008,giannantonio/porciani:2010,schmidt/kamionkowski:2010,desjacques/etal:2011}, i.e.
\begin{equation}
\label{eq:deltag}
\delta_g(\vk,z) = b_1(z) \delta_m(\vk,z) + \bphi(z) \fnl\phi(\vk)\;.
\end{equation}
Here, $\vk$ is the wavenumber and $\delta_m$ is the evolved matter density contrast. While the linear bias parameter $b_1$ arises also when the primordial curvature perturbation is purely Gaussian \citep{kaiser:1984,bardeen/bond/etal:1986,coloe/kaiser:1989,mo/white:1996,sheth/tormen:1999}, the non-Gaussian bias $\bphi$ contributes to the observed galaxy clustering only if $\fnl\ne 0$.
The value of $b_1$ can be extracted from the observed galaxy power spectrum solely by measuring $\delta_g(\vk,z)$ down to the mildly non-linear scales where the 1-loop terms proportional to $b_1$ contribute \cite[see][for a review about galaxy bias]{biasreview}. Model-independent analyses based on the consistency relations of large scale structure \citep{peloso/pietroni:2013,kehagias/riotto:2013,creminelli/etal:2013,peloso/pietroni:2014} offer a complementary approach to robustly measure the large scale bias~\citep{marinucci1,marinucci2}. 

The local mapping Eq.~(\ref{eq:png}) translates into a modulation of the small-scale primordial power spectrum $P_\phi(k_s)$ by long-wavelength perturbations $\phi_l(x)$ with $k_l\ll k_s$. 
A peak-background split argument shows that $\bphi$ is then given by the response of the galaxy number density $n_g$ to a change in the primordial scalar amplitude $A_s$ or, equivalently, to the normalisation amplitude $\sigma_8$ \citep{slosar/etal:2008}, 
\begin{equation}
    \label{eq:pbs}
    \bphi = 2\frac{\partial\ln\bar n_g}{\partial\ln\sigma_8} \;.
\end{equation}
The primordial potential and the matter density contrast are related through $\delta_m(\vk,z) = \mathcal{M}(k,z)\phi(\vk)$ with $\mathcal{M}(k,z) = (2/3)k^2T_m(k)D_{\rm md}(z)/(\Omega_{m}H_0^2)$. Here, $D_{\rm md}(z)$ is the linear growth rate normalized to $a = (1+z)^{-1}$, $T_m(k)$ is the matter transfer function, $\Omega_m$ the present-day mean matter density and $H_0$ the Hubble constant. Eq.~(\ref{eq:deltag}) makes clear that the scale-dependent bias constrains the product $\fnl\bphi$, and not just $\fnl$.
\begin{figure}
    \centering
    {\includegraphics[height = 0.54\textheight, angle = 180] 
    {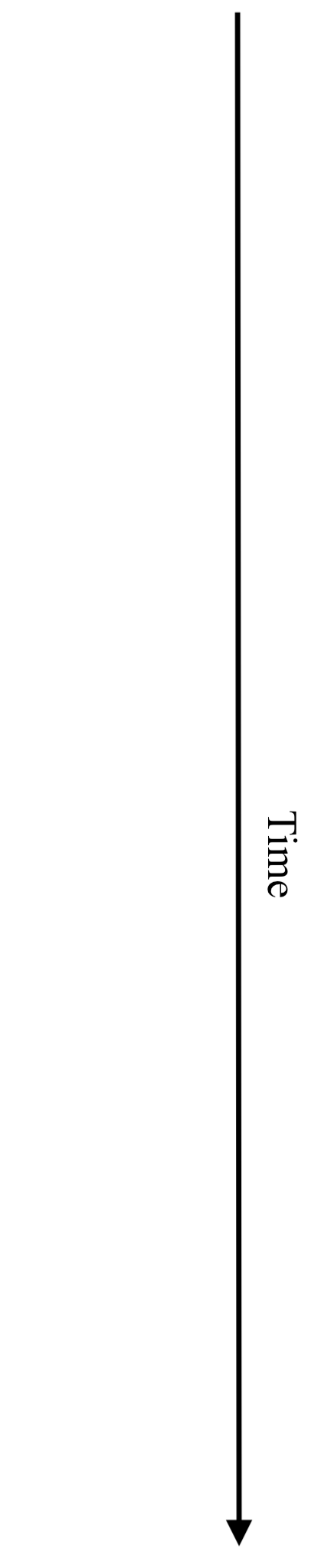}}
    {\hspace{-1.2cm}\includegraphics[height = 0.54\textheight, angle = 180] 
    {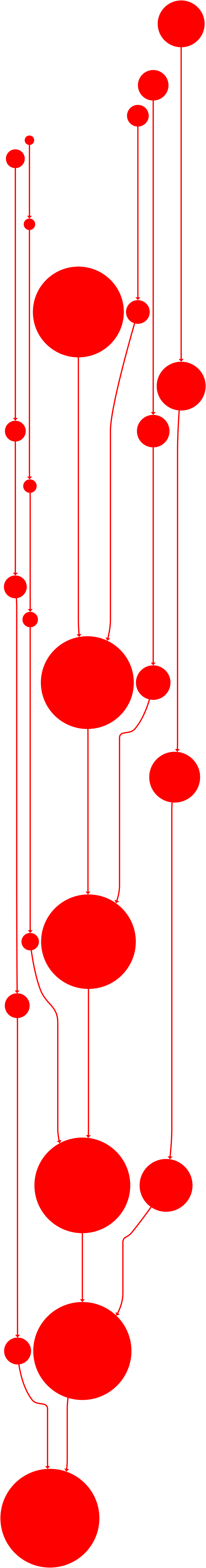}}
    {\hspace{-1.765cm}{\includegraphics[height = 0.55\textheight, angle = 180] 
    {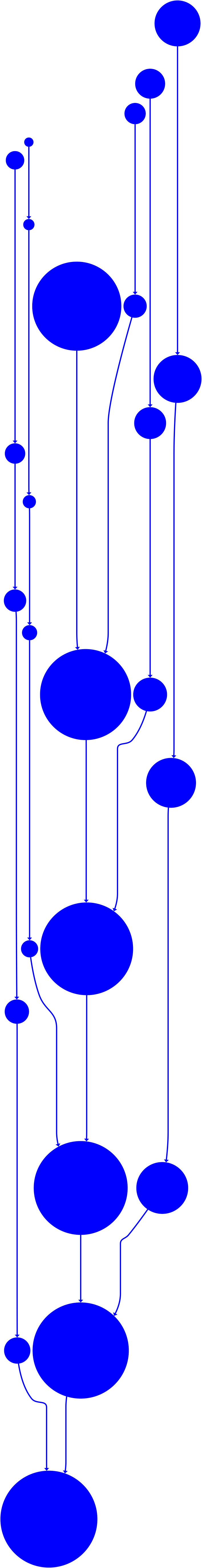}}}
    \caption{
    A change in the normalization amplitude $\sigma_8$ affects halo/subhalo masses, merging time etc. and produces smooth variations in the structure of the trees generated by \Gal. In particular, increasing $\sigma_8$ (at fixed final halo mass, $M_h = 10^{12} \mathrm{M}_\odot$) leads to an earlier merging of the subhalos and, thereby, increases the mass of the parent halo throughout its assembly history as illustrated in the figure. We show two trees run with $\sigma_8 = 0.81$ (red) and $\sigma_8 = 0.83$ (blue). Each circle represents a halo/subhalo and its size is proportional to the mass; the length of the lines is proportional to the merging epoch.}
    \label{fig:trees8}
\end{figure}

The clustering of dark matter halos is imprinted in the galaxy bias parameters. In particular, the non-Gaussian bias of dark matter halos takes the simple form $b_\phi^h(M_h,z) = 2\delta_c(b_1^h(M_h,z) - 1)$ for a universal halo mass function, where $b_1^h$ is the linear halo bias and $\delta_c\approx 1.683$ is the critical threshold for (spherical) collapse. This relation is usually used within LSS analysis with the EFTofLSS~\cite{} to constrain the amplitude of primordial non-gaussianities, which for local type are parametrized by $f_{\rm NL}$. It was shown~\cite{reid/etal:2010,lazeyras/etal:2022} that the relation $b_\phi(b_1) = 2\delta_c (b_1-p)$, with $p = 1$ for halos, is not a good description of the PNG bias of halos and galaxies selected by other properties beyond the halo mass, like the stellar mass $M_\star$. 

\subsection{Modelling galaxy assembly bias}

In the Halo Occupation Distribution (HOD) approach \cite[see for instance][]{benson/etal:2000,scoccimarro/etal:2001,berlind/weinberg:2002,kravtsov/etal:2004,zheng/etal:2005}, galaxy abundances are determined by the mass function $\bar n_h(M,z)$ of host dark matter halos and by the Halo Occupation Distribution (HOD). Therefore, as emphasized in \cite{voivodic/barreira:2021}, galaxy bias parameters can be computed from the change (or response) of $\bar n_h(M,z)$ and the HOD to some long-wavelength perturbation. 

In plain words, let
\begin{equation}
    n_g(X,z) = \int\! dM_h\,\bar n_h(M_h,z)\, \big[N_c(X|M_h,z) + N_s(X|M_h,z)\big]
\end{equation}
be the comoving number density of galaxies $n_g(X,z)$ at redshift $z$ with observed property $X$. 
Here, $N_c(X|M_h,z)$ (resp. $N_s(X|M_h,z)$) is the (average) number of central (satellite) galaxies with a given property $X$ residing in halos of mass $M_h$ at redshift $z$. These conditional means can be further decomposed into
\begin{equation}
    N_{c,s}(X|M_h,z) = \bar N_{c,s}(M_h,z)\, P_{c,s}(X|M_h,z)
\end{equation}
where $\bar N_c(M_h,z)$ (resp. $\bar N_c(M_h,z)$) is the average number density of central (satellite) galaxies per halo, and $P_c(X|M_h,z)$ is the probability distribution function (PDF) of $X$ conditioned to a halo mass $M_h$ and redshift $z$ (and likewise for satellites galaxies).

The non-Gaussian bias $\bphi$ of this galaxy sample is given by
\begin{align}
    \label{eq:galaxyPNG}
    \bphi(X,z) &= 2\frac{\partial\ln n_g}{\partial\ln\sigma_8}(X,z) \\
    &=\bar{b}_\phi(X,z) + \Delta\bphi(X,z) \nonumber \;,
\end{align}
where
\begin{align}
\label{eq:barbphi}
    \bar{b}_\phi(X,z) = \frac{1}{\bar n_g}&\int\!dM_h\, \bphi^h(M_h,z)\, \bar n_h(M_h,z) \\
    & \times\big[ N_c(X|M_h,z) + N_s(X|M_h,z)\big] \nonumber
\end{align}
is a weighted averaged of the non-Gaussian halo bias (the response of the halo mass function) solely. The non-Gaussian halo bias is $b_\phi^h(M_h,z) = 2\,\partial \ln{\bar{n}_h(M_h,z)}/\partial \ln{\sigma_8}$ by definition. Similarly,
\begin{align}
    \Delta \bphi(X,z) &= \frac{1}{\bar n_g}\int\!dM_h\,
    \big[f_c \Delta\bphi^c(X|M_h,z) + f_s \Delta\bphi^s(X|M_h,z)\big] 
    \nonumber \\
    &\qquad \times \bar n_h(M_h,z)\, \big[N_c(X|M_h,z) + N_s(X|M_h,z)\big] \label{eq:assbphi}
\end{align}
is the contribution arising solely from the response of the HOD to a change in $\sigma_8$. Here, $f_c=f_c(X|M_h,z)$ and $f_s=f_s(X|M_h,z)$ are the fraction of central and satellite galaxies with properties $X$ and residing in halos of mass $M_h$ at redshift $z$, whereas
\begin{align}
    \label{eq:assbR}
    \Delta\bphi^{c,s}(X|M_h,z) &= 2\frac{\partial\ln N_{c,s}}{\partial\ln\sigma_8}(X|M_h,z) \\
    &= 2\frac{\partial\ln\bar N_{c,s}}{\partial\ln\sigma_8}(M_h,z)+2\frac{\partial\ln P_{c,s}}{\partial\ln\sigma_8}(X|M_h,z) \nonumber
\end{align}
is the corresponding non-Gaussian assembly bias. 
$\Delta\bphi^{c,s}$ is the fractional change in the number of central or satellite galaxies as $\sigma_8$ is varied.
Our definitions are related to the response $R_\phi^g(X|M_h,z)$ introduced in \cite{voivodic/barreira:2021} through
\begin{equation}
    R_\phi^g(X|M_h,z) =\Big(\bar N_c\,\Delta\bphi^c+\bar N_s\,\Delta\bphi^s \Big)(X|M_h,z) \;,
\end{equation}
which shows that $R_\phi^g(X|M_h,z)$ is the weighted sum of the central and satellite non-Gaussian assembly bias parameters.

We will focus on the non-Gaussian "assembly" bias $\Delta\bphi^{c,s}(X,z)$ since the contribution $\bar b_\phi$ arising purely from the response of the halo mass function $\bar n_h(M,z)$ can be extracted from pure N-body simulations solely for generic primordial bispectrum shapes \citep[e.g.][]{dalal/etal:2008,grossi/etal:2009,pillepich/etal:2010,tseliakhovich/hirata/slosar:2010,desjacques/seljak:2010,shandera/etal:2011,smith/loverde:2011,desjacques/etal:2011_2,wagner/verde:2012,smith/ferraro/loverde:2012,scoccimarro/etal:2012,biagetti/etal:2017,chan/etal:2019}. 

\subsection{A semi-analytical model of galaxy formation: {\small GALACTICUS}}

Galaxy formation is a complex process involving gas inflow and cooling onto dark matter halos, star formation and self-regulation via feedback from supernovae, AGN etc. \cite[see][reviews on this topic]{mo/vdb/white:2010,Somerville:2014ika}.
To model the process of galaxy formation we make use of the semi-analytic galaxy formation model (SAM) \Gal \citep{galacticus} rather than more realistic but computationally expensive hydrodynamical simulations of the large scale distribution of galaxies. 

The \Gal model is able to generate realistic merger histories for halos of any given mass and redshfit (once given cosmological and power spectrum parameters) following the algorithm of \cite{2008MNRAS.383..557P}, who calibrated their method to match results from cosmological N-body simulations. It then solves the physics of galaxy formation in the resulting merging hierarchy of halos through a combination of differential evolution (to describe processes such as gas cooling, star formation, and feedback), and impulsive events (such as galaxy mergers). This results in realizations of galaxy populations for the required halo masses and redshifts. These predictions include both physical properties (e.g. stellar masses), and observable properties (e.g. broad-band and emission line luminosities). \Gal has previously been used to model the population of emission line galaxies to be studied by the \emph{Roman} telescope \citep{2018MNRAS.474..177M,2019MNRAS.486.5737M,2019MNRAS.490.3667Z,2021MNRAS.501.3490Z,2021MNRAS.505.2784Z}.

The baryonic physics of \Gal is described by 30 model parameters, which have been constrained using a variety of observational datasets\footnote{See \href{https://github.com/galacticusorg/galacticus/wiki/Constraints:-Baryonic-Physics}{here} for the full list of datasets used.}. The model for the H$\alpha$ luminosity functions has been constrained from HiZELS~\citep{2013MNRAS.428.1128S} and GAMA~\citep{2013MNRAS.433.2764G} observations, while the g and r-band luminosity functions of SDSS galaxies from~\cite{2009MNRAS.399.1106M} provide constraints on galaxy colours. 

We run \Gal using the Planck fiducial cosmology~\footnote{The fiducial cosmological parameters are $\{H_0, \Omega_m, \Omega_\Lambda, \Omega_b, n_s\} = \{67.36,0.31530, 0.04930, 0.04930, 0.9649\}$.}~\cite{Planck:2018jri}. Our fiducial normalization amplitude is $\sigma_8^\text{fid}=0.81$.

\subsection{Extracting the non-Gaussian galaxy bias}

To investigate the impact of the galaxy assembly histories on the non-Gaussian assembly bias $\Delta\bphi(X,z)$ with \Gal, we choose a final halo mass $M_h$ and redshift $z$ and generate $N_h$ Monte-Carlo mass assembly histories for a few different values of $\sigma_8$. This allows us to (numerically) directly measure the responses $\frac{\partial\ln N_c}{\partial\ln\sigma_8}$ and $\frac{\partial\ln N_s}{\partial\ln\sigma_8}$ for a given $M_h$ and $z$. We will consider a single redshift, $z=1$, but repeat this procedure for a few different values of $M_h$ in the range $[3\times10^{10},10^{12}]\, \mathrm{M}_\odot$, which brackets the characteristic mass $M_*(z=1)\sim 10^{11}\ \mathrm{M}_\odot$ of halos virializing at redshift $z=1$ in our fiducial cosmology.



We implement the peak-background split expectation eq.~(\ref{eq:assbR}) separately for central and satellite galaxies and compute $\Delta\bphi^{c,s}(X|M_h,z)$ from the random realizations of the merger trees as follows:
\begin{equation}
    \label{eq:responseRc}
    \Delta\bphi^{c,s}(X|M_h,z) = \frac{1}{|\delta_{\sigma_8}|}\left[\frac{N_{c,s}^\text{high}(X|M_h,z) - N_{c,s}^\text{low}(X|M_h,z)}{N_{c,s}^\text{fid}(X|M_h,z)}\right] 
\end{equation}
where $N_c^\text{fid}$ , $N_c^\text{high}$ and $N_c^\text{low}$ are the total number of central galaxies obtained from $N_h$ realizations of halos with mass $M_h$ and redshift $z$ in the fiducial cosmology, and in two "separate" universes with slightly different normalisation amplitudes: $\sigma_8^\text{high}~=~0.83$ and $\sigma_8^\text{low}~=~0.79$. Hence, the fractional change $\delta_{\sigma_8}~=~(\sigma_8^\text{high}-\sigma_8^\text{fid})/\sigma_8^\text{fid}$ in the normalization amplitude is $\simeq 2.5$\%. 

To reduce as much as possible the fluctuations due to shot noise, we run $N_h = 10^5$ trees for the halo masses $M_h = \{3\times 10^{10},10^{11}\} \mathrm{M}_\odot$ and $N_h = 10^{4}$ for $M_h = \{3\times 10^{11},10^{12}\} \mathrm{M}_\odot$. The mass resolution for each run is fixed to the minimum value $M_h = 5\times 10^{9} \mathrm{M}_\odot$ since galaxies living in halos with smaller masses are faint and not easily observed due to their low stellar content. To study the convergence of the trees in our case of study, we have run $N_r = 10$ realizations with different random seeds. The scatter among the different realizations was used to estimate the errorbars for our measurements. They vary significantly among panels due to different choices of binning and changes in the relative fraction of central and satellite galaxies as a function of halo mass.

The primary galaxy observables $X$ we are interested in are the galaxy colours defined by the $(g, r, i,\dots)$ filters, and the emission line strengths such as the H$\alpha$ luminosity $L_\alpha$. They are among the basic quantities directly measured in optical or in emission line galaxy surveys. For instance, the {\small BOSS}~\footnote{\url{https://www.sdss3.org/}} galaxy samples ({\small LOWZ} and {\small CMASS}) are constructed using colour-magnitude cuts~\citep{BOSS_sel}, whereas the {\small Euclid}~\footnote{\url{https://sci.esa.int/web/euclid}} and {\small WFIRST}~\footnote{\url{https://www.jpl.nasa.gov/missions/the-nancy-grace-roman-space-telescope}} galaxy samples will be constructed from the measured H$\alpha$ line fluxes \citep{Euclid_sel,Roman_sel}. Furthermore, {\small SPHEREx}~\footnote{\url{https://www.jpl.nasa.gov/missions/spherex}} will measure molecular and PAH emission in the $\mu m$ range~\citep{SPHEREx_sel}. 

We shall also explore the dependence of $\Delta\bphi$ on secondary properties such as the stellar mass $M_\star$ of the simulated galaxies, the black hole mass/accretion rate $M_\text{BH}$, $\dot M_\text{BH}$ and morphological measures such as the bulge-to-disk ratio $s = R_\text{sph}/R_\text{disk}$ (where $R_\text{sph}$ and $R_\text{disk}$ are the size of the bulge and the disk respectively). Reliable morphology measurements cannot be obtained at high redshift $z\gtrsim 1$ from current state-of-the-art imaging data. Our results for some secondary observables are summarized in Appendix \S\ref{app:secondary}.


\section{Results}
\label{sec:results}

In this Section, we present results for the dependence of the non-Gaussian assembly bias $\Delta\bphi^{c,s}$ on different galaxy properties. We will focus on the $g-r$ colour \cite[used to distinguish between red and blue galaxies, e.g.][]{bell/etal:2004} and the H$\alpha$ line luminosity \citep[which is a prime indicator of star formation, e.g.][]{kennicutt:1983}. Studying the dependence of the non-Gaussian $\bphi$ on these observables is of much interest for an optimal, well motivated choice of priors. 

\subsection{Galaxies selected by colour magnitude}

\begin{figure}
    \centering
    \includegraphics[width = 0.49\textwidth]{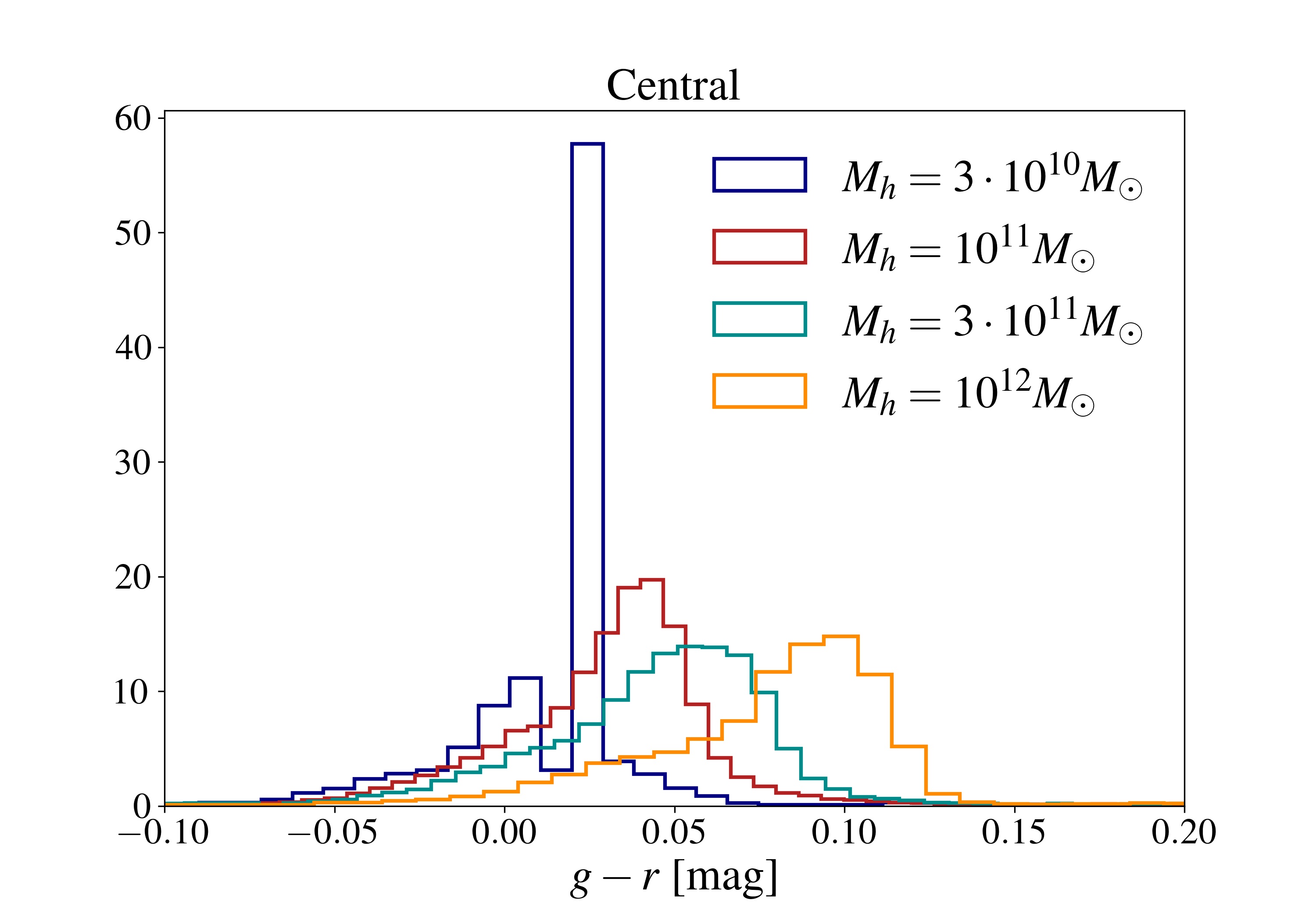}
    \includegraphics[width = 0.49\textwidth]{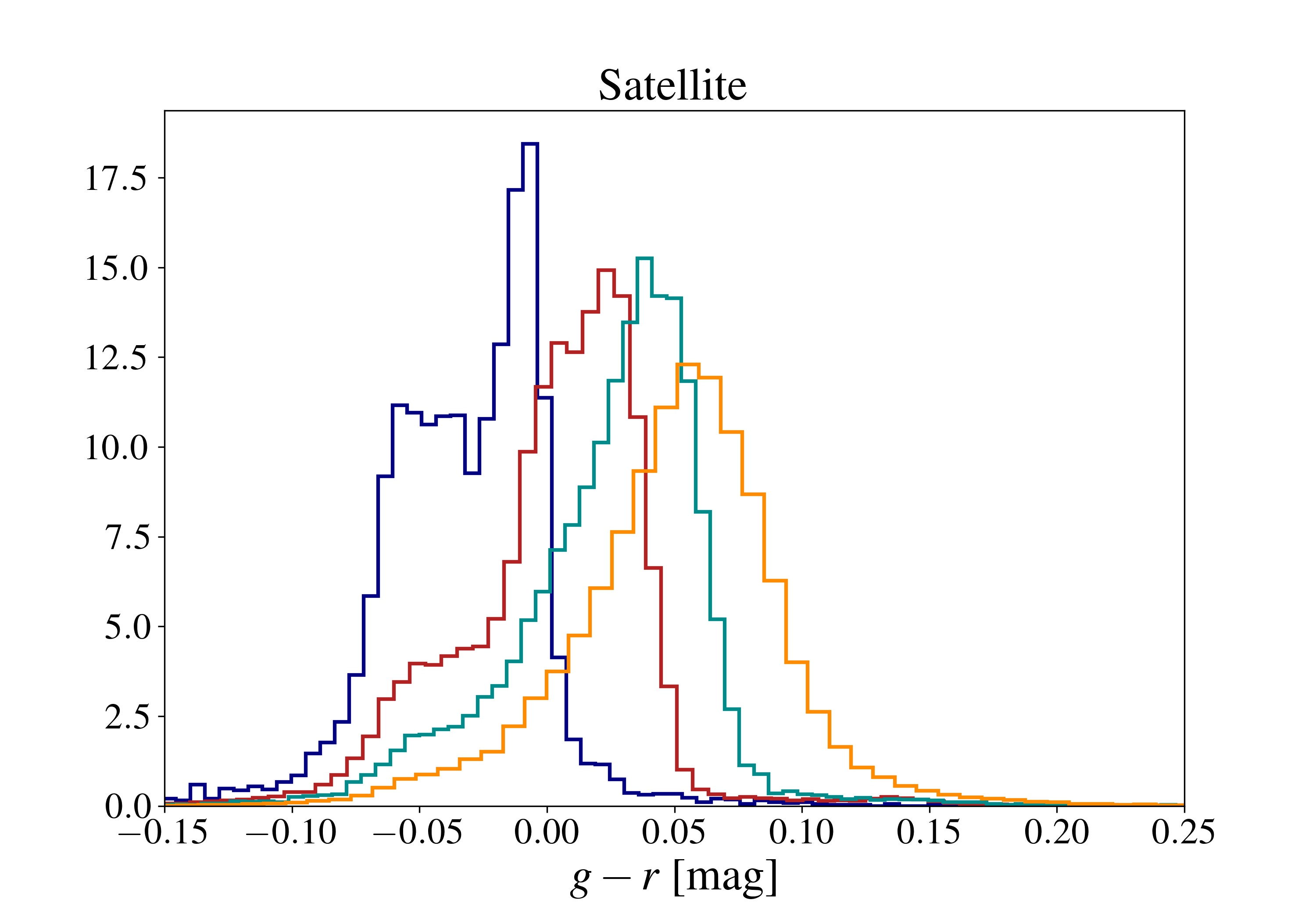}
    \caption{Probability density of $z=1$ galaxies as a function of the colour $g-r$ for the various halo masses $M_h$ considered in our analysis. Galaxies with $g-r\gtrsim 0.5$ are not produced in significant numbers for the range of halo masses considered here (see text).}
    \label{fig:pdf_color}
\end{figure}

Fig.~\ref{fig:pdf_color} shows the $g-r$ colour distribution at redshift $z=1$ for the range of halo mass considered here. The color distribution of \Gal was studied within the {\small CosmoDC2} project, see~\cite{CosmoDC2}.
The lack of relatively "red" galaxies with $g-r\gtrsim 0.5$ originates from the fact that these are either satellite galaxies in high mass halos $M_h\gtrsim 10^{13}\, \mathrm{M}_\odot$, the hot circum-galactic medium of which has cut-off the supply of new gas so that satellites no longer form stars, or galaxies residing in very low mass halos $M_h\lesssim 10^9\ \mathrm{M}_\odot$ that are unable to accrete gas (due to their shallow potential wells). While the latter have very low stellar content and would likely fall below the detection limit of realistic galaxy surveys, the former are detected by current galaxy surveys, albeit in small numbers since their massive host halos are rare. The halo mass range considered here and, thereby, the distributions shown in Fig.~\ref{fig:pdf_color} sample the bulk of the $g-r$ colour distribution at $z=1$. The difference seen with \cite{barreira/TNG:2020}, who report measurements for galaxies with colour magnitude as large as $g-r\sim 1$, presumably arises because their analysis includes halos with masses up to $M_h\sim 5\times 10^{14}\ \mathrm{M}_\odot$ which we do not simulate. Moreover, the lack of galaxies, observable in fig.~\ref{fig:pdf_color}, with $g-r \gtrsim 0.2$ could be the result of intrinsic differences between galaxy formation models used in SAMs codes and hydro-dynamical simulations.

\begin{figure}
    \centering
    \includegraphics[width = 0.48\textwidth]{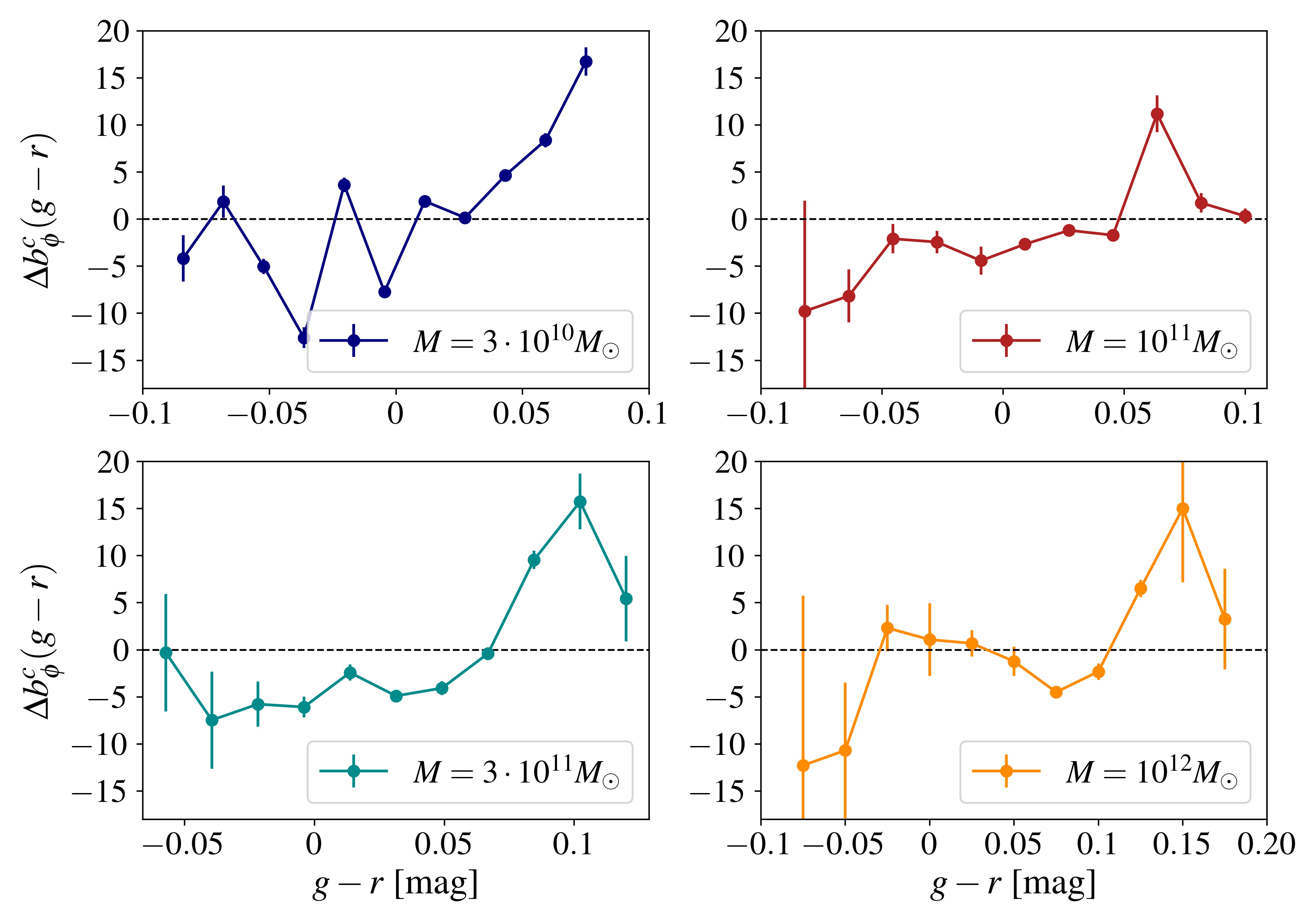}
    \includegraphics[width = 0.47\textwidth]{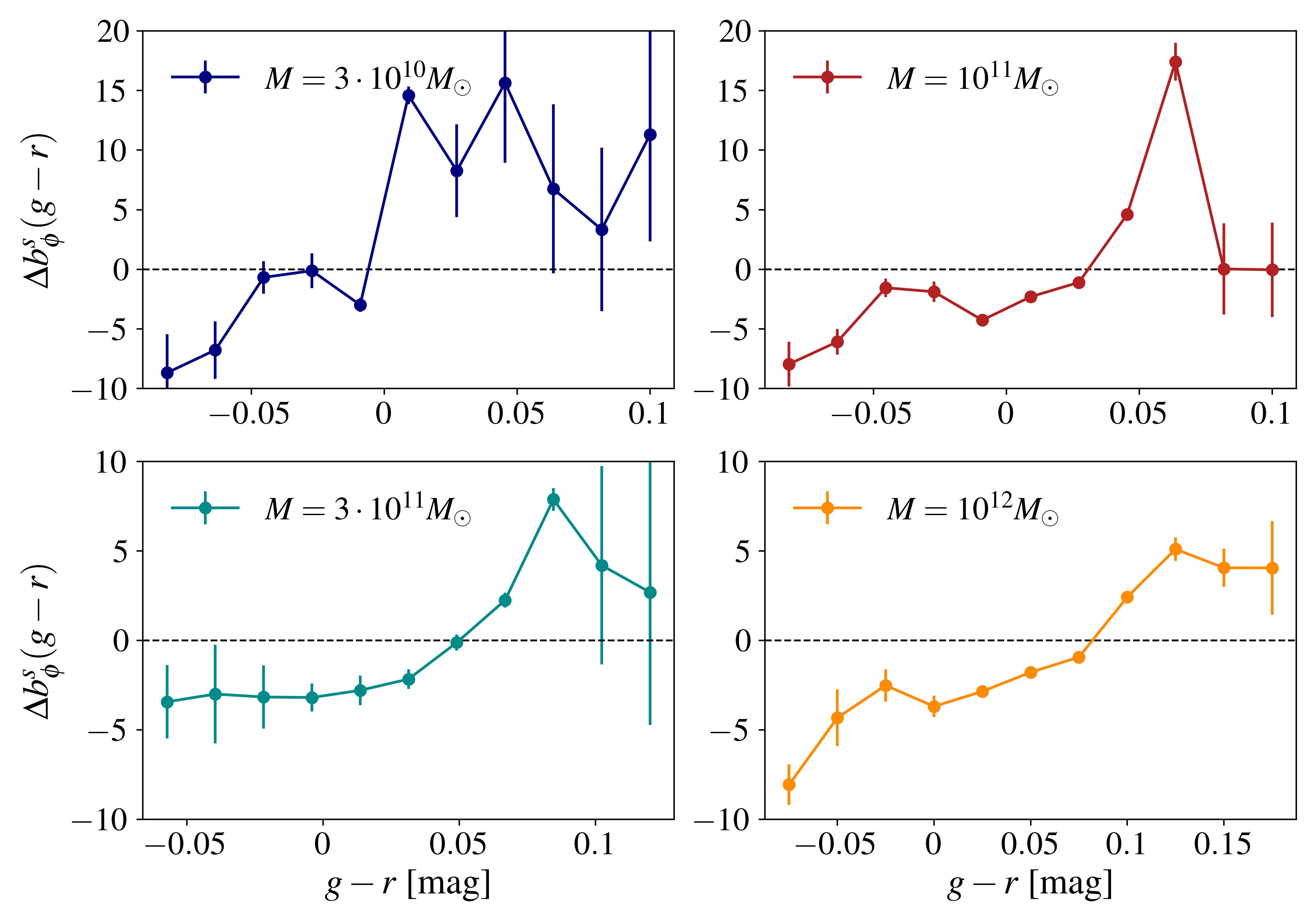}
    \caption{Non-Gaussian assembly bias $\Delta\bphi^{c,s}$ for galaxies selected by their $g-r$ colour magnitude. Results are shown for central (upper panels) and satellite (lower panels) galaxies as a function of the host halo mass $M_h$. The errorbars are computed from the scatter between random realizations of $N_h=10^5,10^4$ merger trees for $M_h = \{3\times10^{10},10^{11}\} \mathrm{M}_\odot$ and $M_h = \{3\times10^{11},10^{12}\} \mathrm{M}_\odot$ respectively.}
    \label{fig:bphiColor}
\end{figure}

Fig.~\ref{fig:bphiColor} displays the non-Gaussian assembly bias $\Delta\bphi^{c,s}(g-r)$ of $z=1$ central and satellite galaxies for different $M_h$ as indicated in the figure. We find that $\Delta\bphi^{c,s}$ is positive (negative) when $g - r$ is larger (smaller) than the mean colour of the sample. This trend originates from the fact that an increase in $\sigma_8$ results, on average, in a faster mass assembly history. Consequently, the final dark matter halo hosts a higher number of older, "red" galaxies and, correspondingly, a smaller number of younger ones. 
This effect is visible both for central and satellite galaxies. Furthermore, in light of the findings of \cite{reid/etal:2010}, we expect it to be correlated with a change in the average formation redshift of the host dark matter halos.

Overall, the non-Gaussian assembly bias can change by up to a factor of $\mathcal{O}(10)$ depending on the colour, in agreement with the results of \cite{barreira/TNG:2020} extracted from detailed hydro-dynamical simulations (of a different galaxy formation model). Furthermore, in most panels there is a prominent peak at $g-r\sim 0 - 0.1$ on top of the broad tilt produced by the change in merging histories of dark matter halos and galaxies. The position of the peak matches the position of the maximum of $P_{c,s}(g-r|M_h,z)$, around which $\Delta\bphi^{c,s}$ changes rather abruptly. This effect is partially erased when larger $g-r$ bins are adopted. There are other features which appear to be robust to statistical uncertainties. Although they might be caused by some feature in the stellar population spectra, their origin is unclear. 


It is instructive to contrast these measurements to the contribution $\bar{b}_\phi$ given by Eq.~(\ref{eq:barbphi}), which arises from the halo clustering solely. Assuming the halo mass function of ref.~\cite{Tinker:2010my} and the universality relation $\bar{b}_\phi=2\delta_c(b_1-1)$, we find $\bar{b}_\phi(M_h,z=1) = [-0.57, -0.24,  0.20,  0.95]$ for the halo masses $M_h = [3\times10^{10}, 10^{11}, 3\times10^{11}, 10^{12}] \mathrm{M}_\odot$ considered here. In other words, the non-Gaussian assembly bias is significantly larger than the pure "halo" contribution $\bar{b}_\phi$. Although a realistic prediction would have to integrate $\Delta\bphi$ across the range of redshift and halo masses probed by the galaxy survey under consideration, these results confirm again that the assumption of a universality relation $\bphi=\bar{b}_\phi(b_1)$ can be a poor fit for galaxies selected by colour magnitude \cite[see][]{barreira/TNG:2020,Barreira:2021dpt,barreira:2022a}. Furthermore, suitable colour cuts could be applied in order to maximize the information that can be extracted on $\fnl$ from the non-Gaussian bias~\citep{barreira&krause}. For a host halo mass $M_h = 3\times 10^{11} \mathrm{M}_\odot$ for instance, a higher signal-to-noise for the non-Gaussian bias $\fnl\bphi$ is obtained for galaxies with a colour magnitude $g-r\simeq 0.1$ close to the peak of $\Delta\bphi^{c,s}$ although, in practice, the peak of the response is broadened by the range of halo mass and redshift probed by the survey. 

Finally, we have also checked that, at least for host halos of mass $M=3\times 10^{10}\ M_\odot$, the non-Gaussian assembly bias $\Delta\bphi^{c,s}(r-i)$ at fixed $r-i$ color magnitude is similar to $\Delta\bphi^{c,s}(g-r)$ shown here.



\subsection{Galaxies selected by H$\alpha$ luminosity}

\begin{figure}
    \centering
    \includegraphics[width=0.48\textwidth]{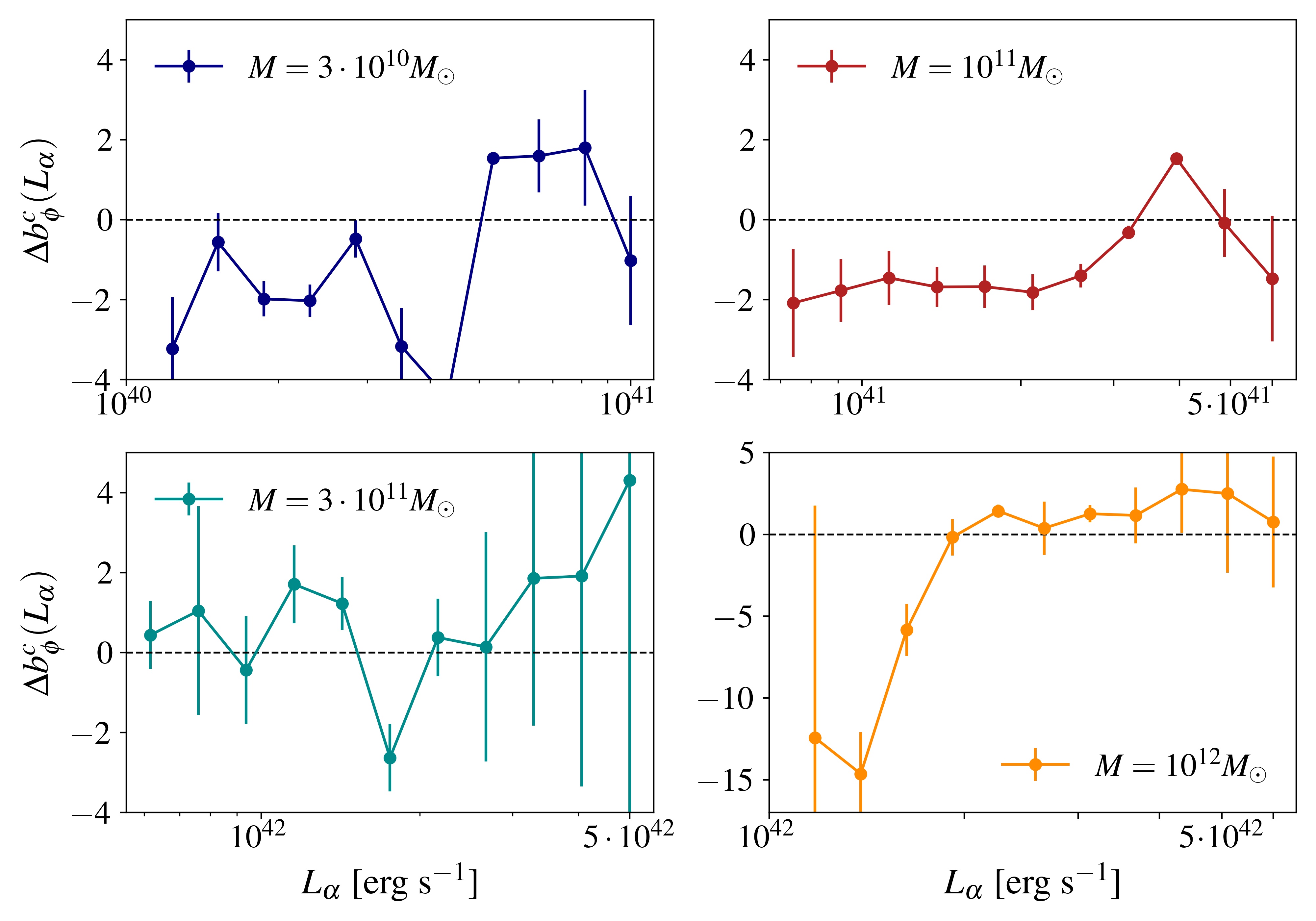}
    \includegraphics[width=0.48\textwidth]{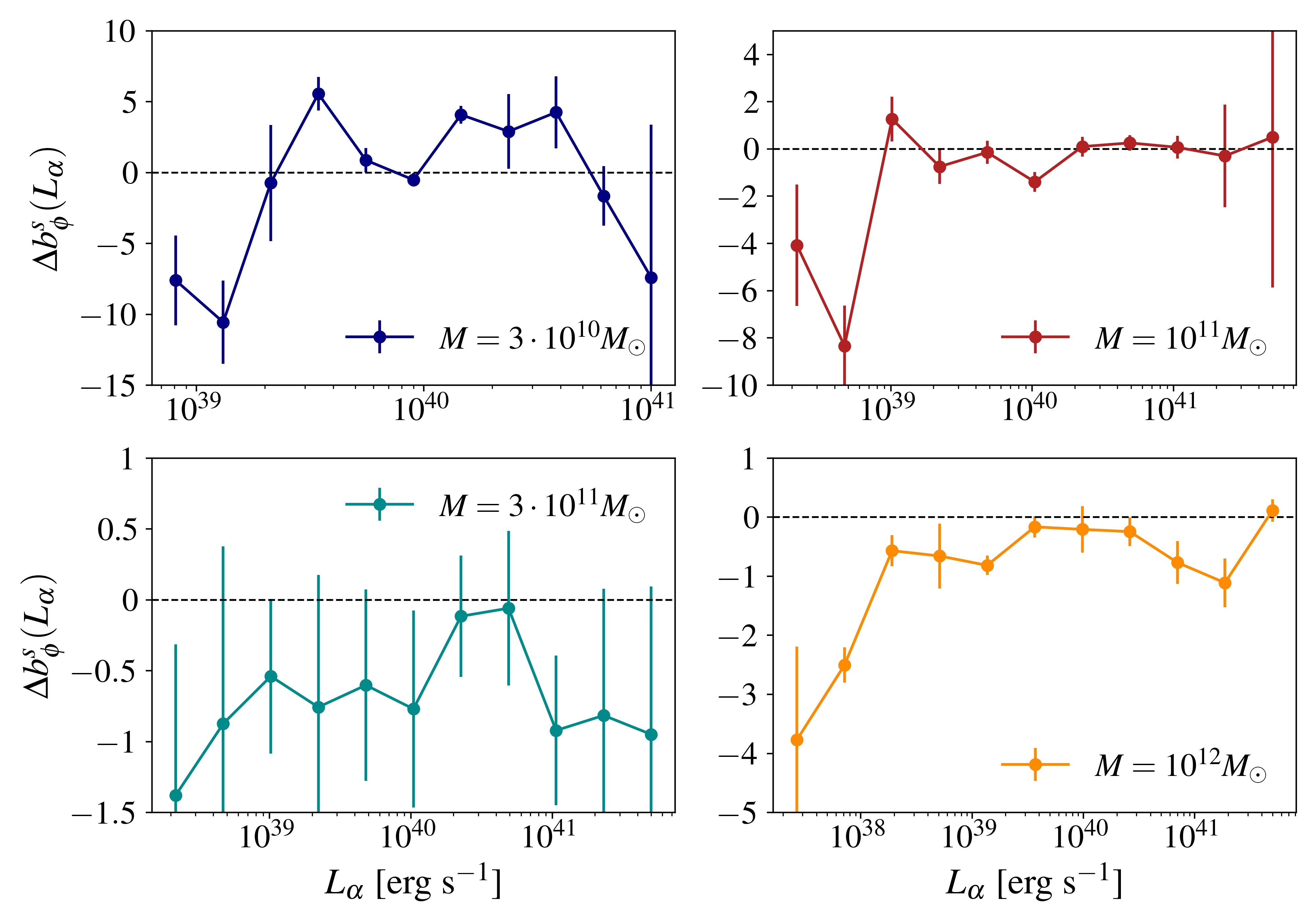}
    \caption{Same as Fig.~\ref{fig:bphiColor}, for the H$\alpha$ luminosity $L_\alpha$. Note that the range of H$\alpha$ luminosity of the mock galaxies increases significantly with the mass of the host halo.}
    \label{fig:bphiEL}
\end{figure}

Fig.~\ref{fig:bphiEL} summarizes our findings for the non-Gaussian assembly bias $\Delta\bphi(L_\alpha)$ as a function of the H$\alpha$ luminosity. For this purpose, \Gal was calibrated against simulations and observations, see~\cite{Zhai:2019hjt} for details. The H$\alpha$ luminosity is calculated at the selected redshift and post-processed in order to include correction from dust extinction in the target galaxy. For a survey such as {\small Euclid}, a H$\alpha$ line flux limit of $2\times 10^{16}\flux$ \citep{EuclidWideSurvey} corresponds to a minimum H$\alpha$ luminosity of $\sim 4\times 10^{41}\lum$ at redshift $z=1$, which is barely reached in our sample of satellite galaxies. 

\begin{figure*}
    \centering
    \includegraphics[width = 0.3\textwidth]{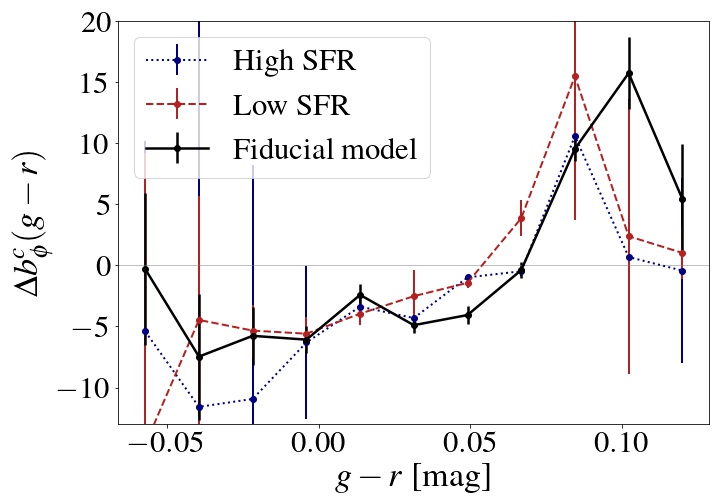}\includegraphics[width = 0.3\textwidth]{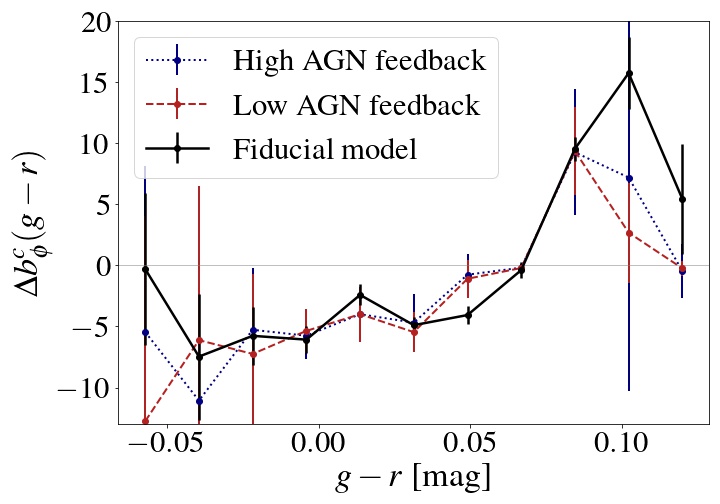}
    \includegraphics[width = 0.3\textwidth]{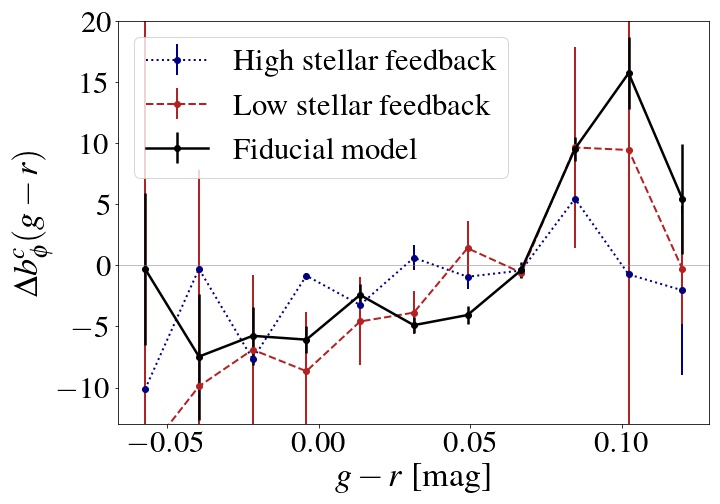}
    \includegraphics[width = 0.3\textwidth]{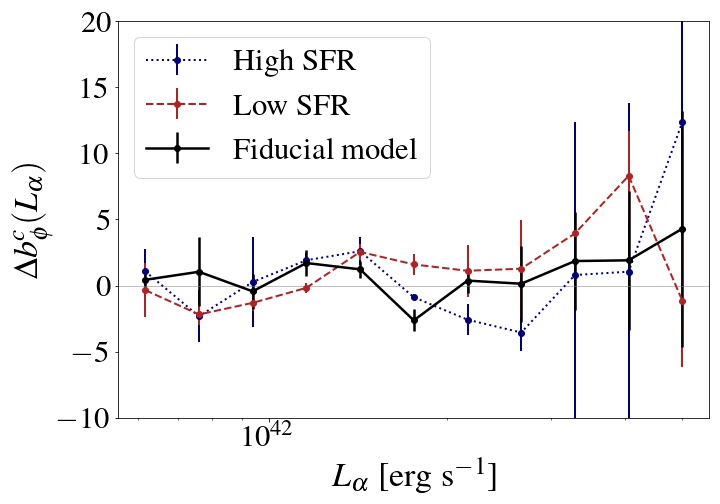}\includegraphics[width = 0.3\textwidth]{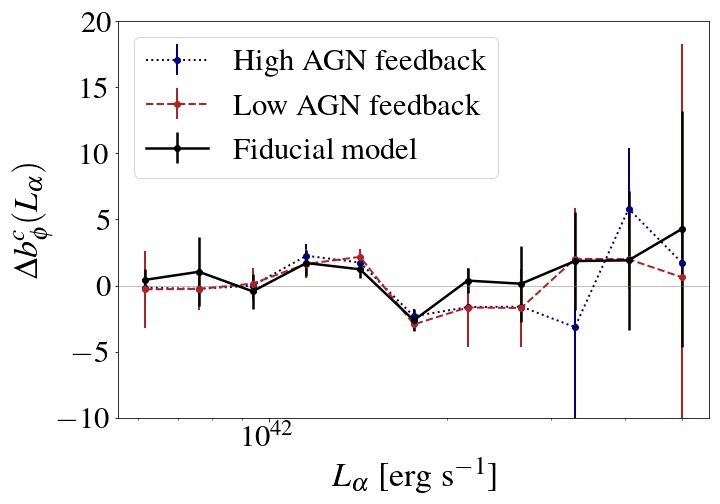}
    \includegraphics[width = 0.3\textwidth]{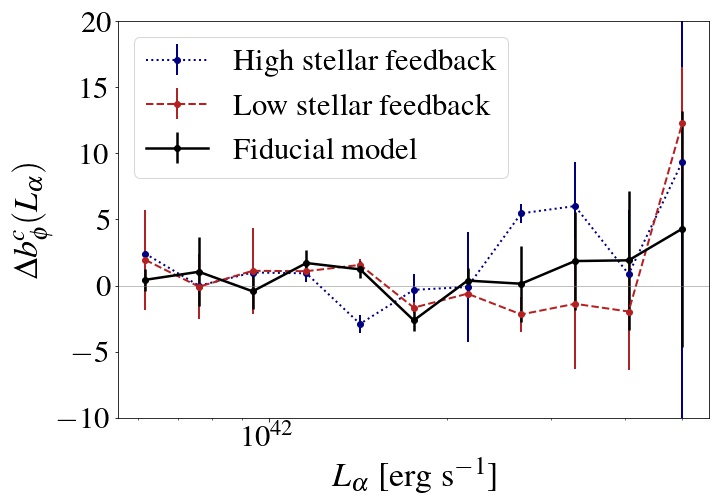}
    \caption{Effect of varying the galaxy formation model on the non-Gaussian assembly bias $\Delta\bphi$ (see text for details). Results are shown for central galaxies populating $z=1$ halos of mass $M_h = 3\times 10^{11} \mathrm{M}_\odot$ and selected either according to their $g-r$ colour magnitude (top panel), or to their H$\alpha$ luminosity (bottom panels).}
    \label{fig:Col_Gmod}
\end{figure*}

Overall, the assembly bias is much smaller for galaxies selected by H$\alpha$ luminosity than $g-r$ colour magnitude. The reason is that $L_\alpha$ mainly depends on the instantaneous star formation rate $\dot\rho_*(z)$, which turns out to be a weak function of the normalization amplitude $\sigma_8$ at the redshift considered here.
For central galaxies we have that the $L_\alpha$ distribution of galaxies for the redshift and halo masses considered here are narrowly peaked, as can be inferred by the luminosity ranges shown in the upper plot of fig.~\ref{fig:bphiEL}. Unlike colour cuts, we do not detect any clear assembly bias trend, even though the departure from $\Delta\bphi$ of order $\sim O(1)$ seen for, e.g., central galaxies in halos of mass $M_h=3\times 10^{10}$ and $10^{11}\ \mathrm{M}_\odot$ (upper row) appear robust to sampling variance. 

Likewise, there is no indication of assembly bias for the satellite galaxies except, possibly, at the lowest luminosities $L_\alpha\lesssim 10^{38 - 39}\ {\rm erg}\ {\rm s}^{-1}$ probed by the mock galaxy catalogs. 
We have checked that the non-Gaussian assembly bias exhibits a similar behaviour when galaxies are selected by their H$\beta$ luminosity (at least for halos of mass $M=3\times 10^{11}\ M_\odot$). All this suggests that setting $\bphi=\bar{b}_\phi$ for emission line galaxies selected by H$\alpha$ luminosity should be a reasonable approximation for the non-Gaussian bias. This, of course, does not preclude the existence of a Gaussian assembly bias (i.e. at the level of $b_1$ etc.) as reported in \cite{jimenez/etal:2021} for instance. This being said, one should remain cautious and avoid taking these results as definitive until a thorough study based on different galaxy formation models (we vary some of the \Gal parameters in section \S\ref{sec:galmod}) and merger trees probing a wider range of halo mass and redshift is carried out.

\subsection{Changing the galaxy formation model}
\label{sec:galmod}

The flexibility of \Gal allows us to study the dependence of our assembly bias measurements on the choice of galaxy formation model. For this purpose, we consider three variations around the fiducial model adopted by \Gal. The first variation is a change the efficiency of the star formation rate (SFR), which should have an impact on the observed galaxy colour and luminosity. The resulting "low SFR" / "high SFR" models have star formation efficiencies twice as low / large as the fiducial one. All the other model parameters are held fixed to their fiducial value.
For the second and third set of models, we modify the heating efficiency of the black hole component (i.e. the efficiency of the heating from the active galactic nuclei, or AGN feedback), and the mean velocity of the stellar feedback outflows. Both parameters control the amount of energy which is dumped into the surrounding gaseous medium, thereby affecting the production of new stars. Here again, we adopt parameter values twice as large / as small as the fiducial ones. For convenience, the values of the parameters changed to produce our different galaxy formation models are summarized in Appendix \S\ref{app:gal}. These physical parameters are expected to have the highest impact on the observed galaxy colours and H$\alpha$ luminosities, as well as the other galaxy observables presented in Appendix~\ref{app:secondary}.


Fig.~\ref{fig:Col_Gmod} displays the results obtained from 6 samples of $10^4$ merger trees of host halos of mass $M_h = 3\times 10^{11}\mathrm{M}_\odot$ virializing at redshift $z = 1$. They are shown as a function of $g-r$ colour magnitude (top panels) and H$\alpha$ luminosity (bottom panels) for central galaxies solely. While the behaviour of $\Delta\bphi^c(g-r)$ remains mostly unchanged, the peak of the assembly bias can be slightly shifted. The effect is most pronounced when a lower efficiency in the SFR is considered. This is due to the fact that a lower star formation efficiency delays star formation and, thereby, increases the relative fraction of younger stars. As a result, this shifts $\Delta\bphi^c$ toward bluer colours (i.e. large values of $g-r>0$).   
Moreover, fig.~\ref{fig:Col_Gmod} shows that, while variations in the AGN feedback efficiency appear to have a negligible impact, variations in the stellar feedback can affect the slope of $\Delta\bphi^c(g-r)$. For all the variations considered here however, the results are always compatible with the trend seen for the fiducial model. 

Likewise, we do not detect large variations in the predicted $\Delta \bphi^c(L_\alpha)$ when the aforementioned model parameters are varied. However, while most of the departures from the fiducial model prediction are within the error bars, some differences can be observed when the SFR and the stellar feedback are varied, mainly at high H$\alpha$ luminosities. Still, investigating these effect further requires a thorough comparison of different SAMs, which is beyond the scope of this work.


\section{Conclusions}
\label{sec:conclusions}


Detecting or constraining primordial non-gaussianity (PNG) is one of the key goals of present and future galaxy surveys. The current constraints on the local PNG parameter $\fnl$ obtained using galaxy clustering data come mainly from the scale-dependent bias effect in the galaxy power spectrum. At leading order, PNG of the local type give a broadband, scale-dependent contribution $\bphi \fnl/k^2$ to the observed galaxy overdensity, where the non-Gaussian bias $\bphi$ can be derived from a peak-background split argument and associated to a change in the normalization amplitude $\sigma_8$. Good priors on $\bphi$ are necessary to minimize the uncertainty on a measurement of $\fnl$ from galaxy clustering statistics such as the power spectrum and bispectrum.


In this paper, we have studied the dependence of $\bphi$ on survey selection cuts and on the physics of galaxy formation using $z=1$ mock galaxy samples produced with the code \Gal, which implements a particular semi-analytical model (SAM) for galaxy formation~\citep{galacticus}. 
The separate universe approach allows us to directly measure $\bphi$ as the response of the number of galaxies to a variation in the amplitude of the primordial gravitational potential. To measure the non-Gaussian assembly bias $\Delta\bphi$ separately, we have produced merger trees with varying $\sigma_8$ but fixed final halo mass, thereby removing the response of the halo mass function (which can be easily calibrated with N-body simulations). We have focused on measurements of $\Delta\bphi$ for galaxies selected by the $g-r$ colour magnitude and by the intensity of H$\alpha$ emission lines. For $g-r$ cuts, our findings are consistent with the previous study of \cite{barreira/TNG:2020} based on detailed hydro-dynamical simulations, that is, $\Delta\bphi(g-r)$ can reach values as large as 
$|\Delta \bphi|\simeq 10$ -- 20 for galaxies selected by $g-r$ colour magnitude (see fig.~\ref{fig:bphiColor} for instance). This is much larger than the non-Gaussian bias contribution arising from the response of the halo mass function solely, which is $\bar{b}_\phi\simeq O(1)$ for the halo masses and the redshift considered in our study.
This strong assembly bias reflects the large dependence of halo formation times on $\sigma_8$ identified by \cite{reid/etal:2010}: at fixed final halo mass, higher values of $\sigma_8$ cause an earlier collapse of halos and, thereby, a earlier star formation so that the final galaxies are populated by older (redder) stars. 

We have also measured the non-Gaussian assembly bias for $z=1$ galaxies selected by H$\alpha$ luminosity $L_\alpha$, which is relevant for forthcoming emission line galaxy (ELG) surveys such as {\small Euclid} or {\small SPHEREx}.
Previous studies have focused on the linear bias $b_1$ of ELGs and found it to be insensitive to the star formation rate (SFR) \citep{angulo/etal:2012,nusser/yepes/branchini:2020}.
Unlike a $g-r$ colour selection, we have found a weaker non-Gaussian assembly bias $\Delta\bphi(L_\alpha)$ for the halo masses and the redshift considered here (see fig.~\ref{fig:bphiEL} for instance). 
The reason presumably is the fact that the strength of the H$\alpha$ line is mainly sensitive to the instantaneous SFR, which is weakly affected by a change of $\sigma_8$ for the redshift analyzed here. The validity of these results is, however, restricted to the redshift and halo masses considered here. Extending the analysis to higher redshift (where the impact of $\sigma_8$ on the SFR might be larger) and a wider range of halo mass is left for future work.
On the one hand, a small value of $|\Delta\bphi| \ll |\bar{b}_\phi|$ would be a good news for future emission-line surveys such as {\small Euclid} or {\small SPHEREx} since accurate priors on $\bar{b}_\phi$ would only require (sufficient) knowledge of the halo occupation distributions $N_c(X|M_h,z)$ and $N_s(X|M_h,z)$ of central and satellite galaxies. On the other hand, a weaker assembly dependence would prevent us from taking advantage of selection cuts to find samples with large  $\bphi$ and, thereby, improve $\fnl$ constraints at the level achieved by the multi-tracer implementation of \cite{barreira&krause}.

Furthermore, we have explored the effect of varying some of the galaxy formation parameters on the non-Gaussian assembly bias (see fig.~\ref{fig:Col_Gmod}) to assess the robustness of our measurements of $\Delta\bphi$. We have focused on the model parameters which likely have the largest impact on the galaxy colours and emission line strengths, that is, the efficiency of the star formation rate, and the AGN and stellar feedbacks. For the few variations considered here, we have not detected major changes in $\Delta\bphi$ although, because $|\bar{b}_\phi|$ is small across the halo mass range probed by our merger trees, these variations could still matter for constraints on $\fnl$. Therefore, these findings need to be confirmed with a more exhaustive investigation. In addition to other colours (e.g. $r-i$) and emission lines (e.g. OII), such a study should compare the outcome of different SAMs and validate the results against hydro-dynamical simulations, which are required to quantify the systematics arising from excursion set or N-body merger trees. 
 

\section*{Acknowledgments}

We thank Alex Barreira, Guido D'Amico, Elisabeth Krause, Adi Nusser, Massimo Pietroni and Ravi Sheth for useful discussions and comments on this work. MM and VD acknowledge support by the Israel Science Foundation (ISF) grant No. 2562/20.

\appendix

\section{Secondary observables}
\label{app:secondary}

\begin{figure}
    \centering
    \includegraphics[width=0.48\textwidth]{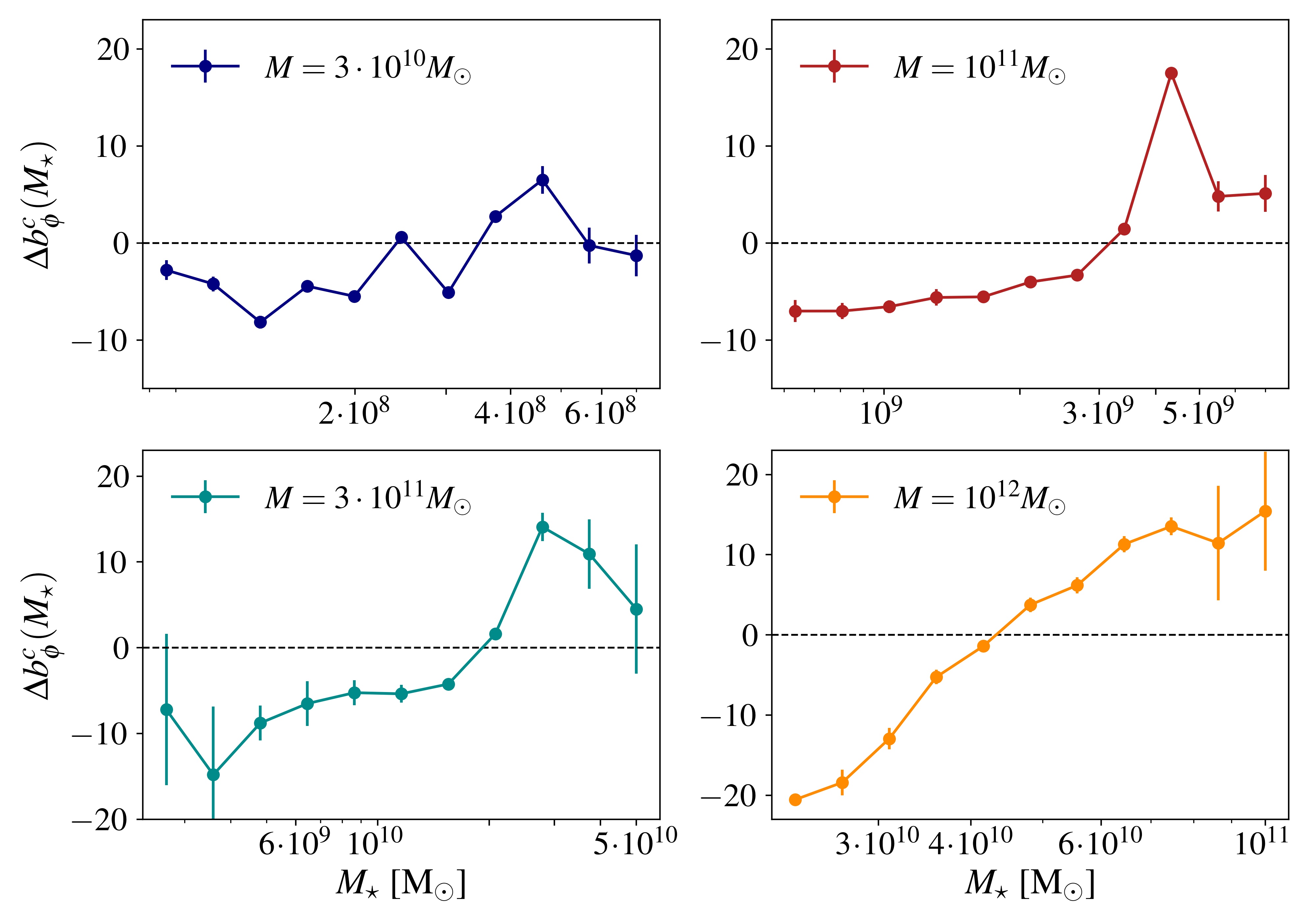}
    \includegraphics[width=0.48\textwidth]{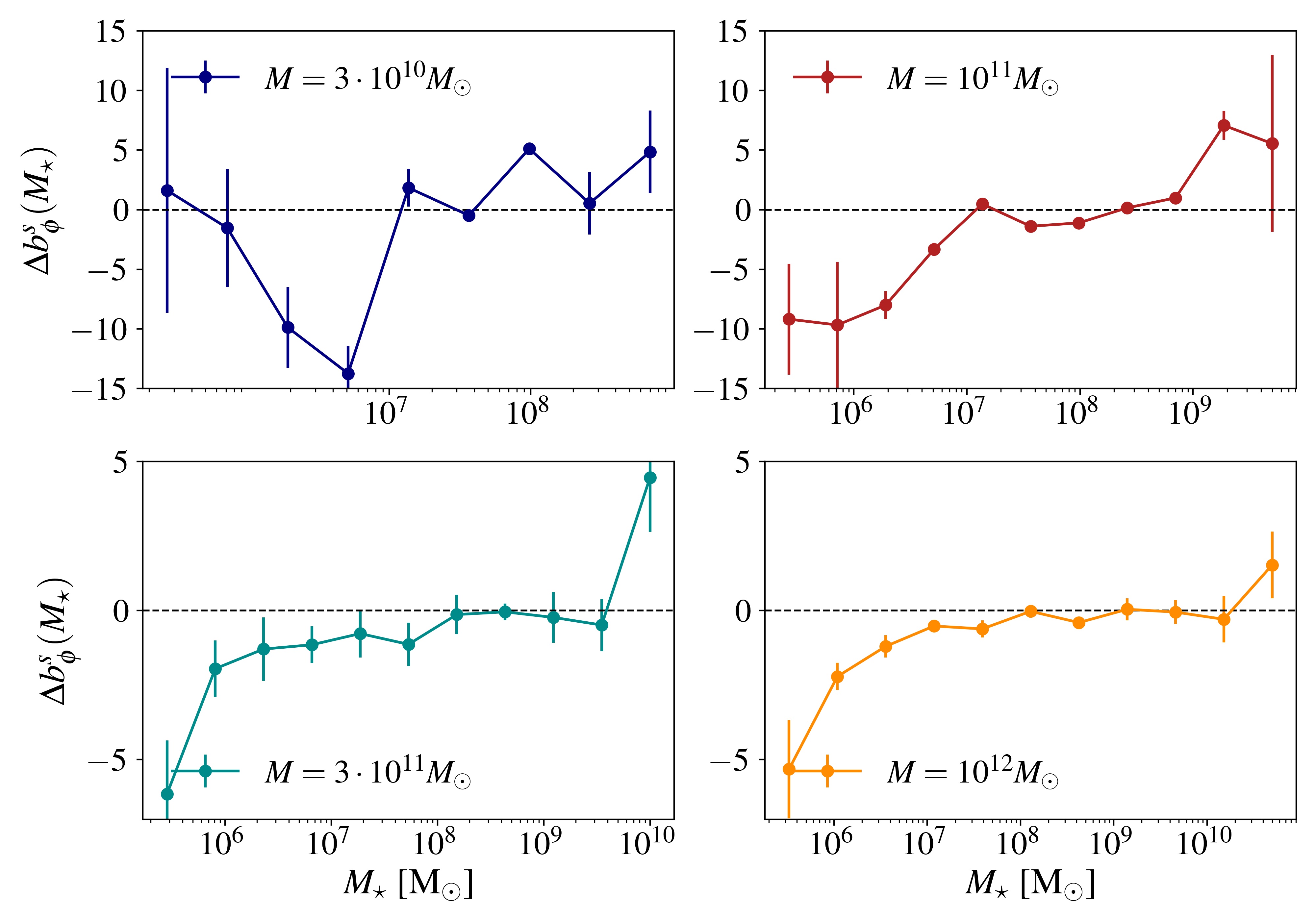}
    \caption{Non-Gaussian assembly bias $\Delta\bphi^{c,s}$ for $z=1$ galaxies selected by stellar mass.}
    \label{fig:bphiMs}
\end{figure}

In this Appendix, we present our measurements of the non-Gaussian assembly bias $\Delta\bphi$ for a few other galaxy properties already considered in literature: the stellar mass $M_\star$, the central black hole mass $M_\text{BH}$ and bulge-to-disk ratio $s$. The galaxy stellar mass is usually inferred from the luminosity through a modelling of the spectral energy distribution (SED). Black hole properties such as mass and accretion rate are correlated with the activity of quasars, which are often used to set observational limits on $\fnl$ as they are highly biased tracers of the underlying matter distribution \citep[see][]{slosar/etal:2008,leistedt/peiris/roth:2014,castorina/etal:2019,mueller/etal:2022}. Furthermore, the current quasar samples cover large comoving volumes at high redshift, thereby reducing the statistical uncertainties. Finally, galaxy morphologies as measured by $s$ can be inferred from imaging surveys, at least at low redshift. 

For galaxies selected by stellar mass, there is a strong assembly bias mainly for central galaxies as is apparent from Fig.~\ref{fig:bphiMs}. The earlier collapse time caused by a higher value of $\sigma_8$ produces galaxies with higher average stellar mass, which explains the measured slope and sign of $\Delta \bphi(M_\star)$.

\begin{figure}
    \centering
    \includegraphics[width=0.48\textwidth]{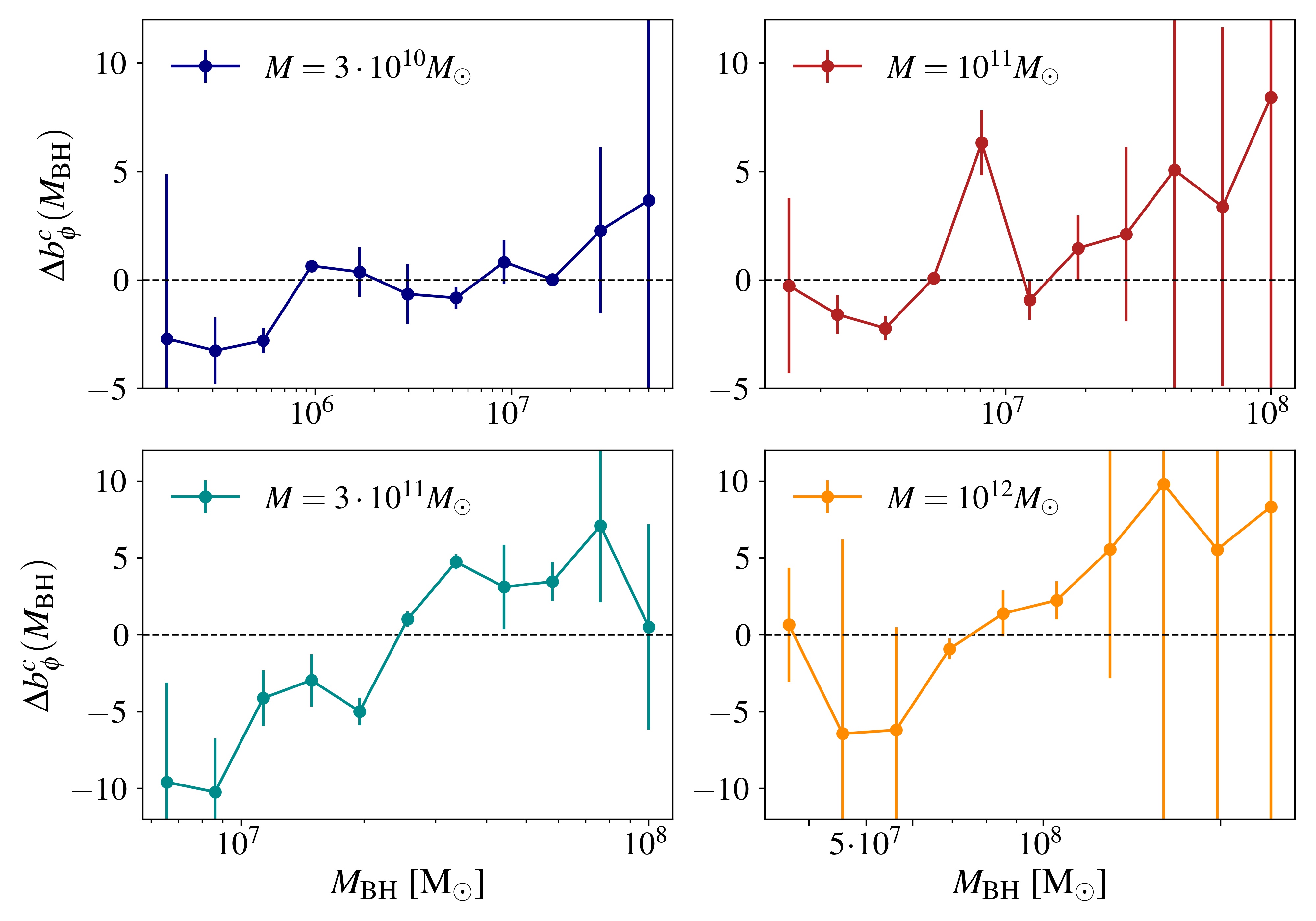}
    \includegraphics[width=0.48\textwidth]{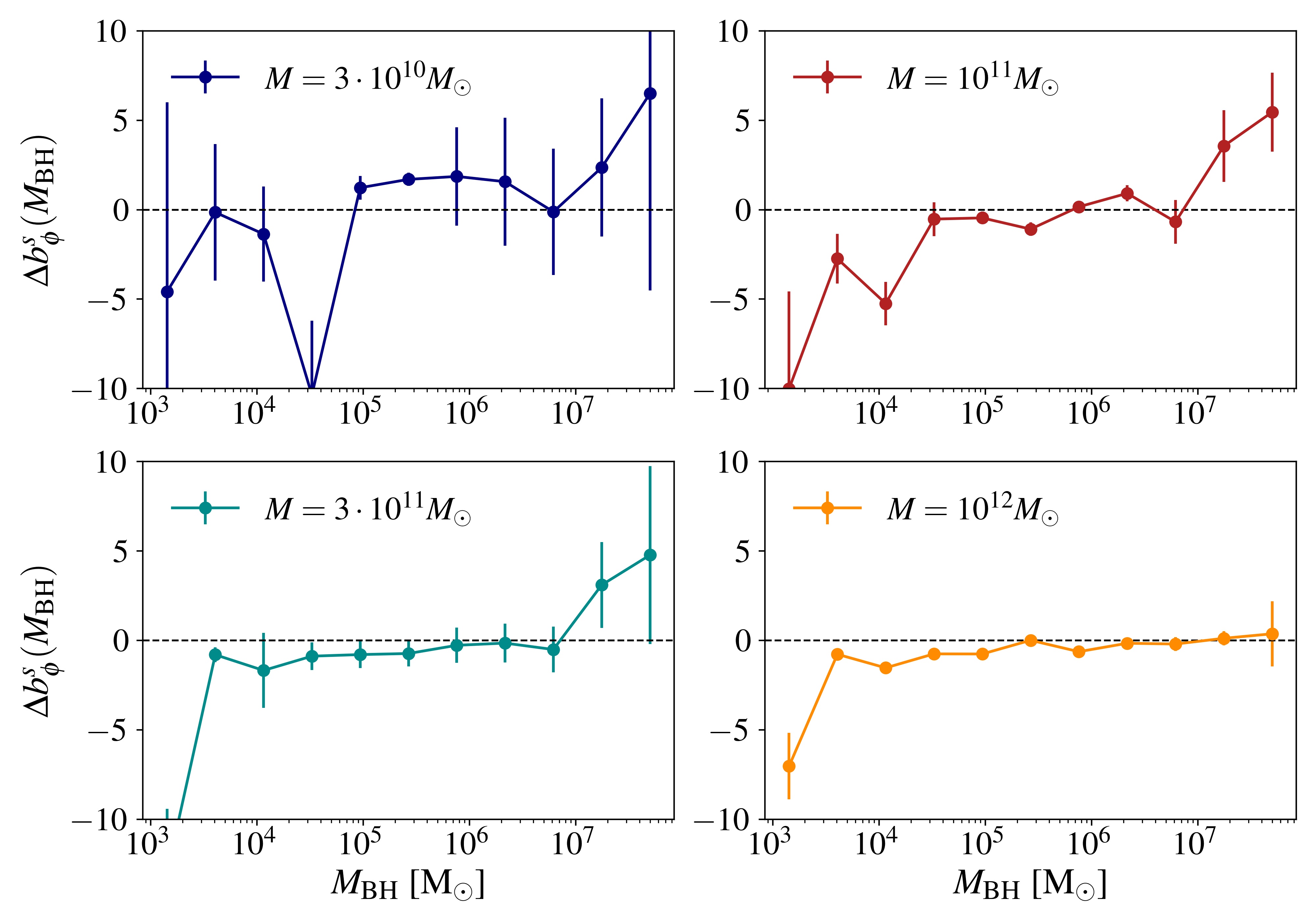}
    \caption{Same as Fig.~\ref{fig:bphiMs} but for galaxies selected by the mass $M_\text{BH}$ of their central black hole.}
    \label{fig:bphiMBH}
\end{figure}

For galaxies selected by the mass of the central black hole, there is a weak transition from negative to positive value of $\Delta\bphi$ for the central galaxies, see~\ref{fig:bphiMBH}. 
This trend, in qualitative agreement with the results of \cite{Barreira:2021BH}, may be explained by a change in the formation time if older galaxies form bigger bulges with larger velocity dispersion and, thereby, larger central black holes (due to the black-hole $M-\sigma$ relation). Galaxies with a central black hole of higher mass $M_{\rm BH}$ present higher average values of $g-r$.

\begin{figure}
    \centering
    \includegraphics[width=0.48\textwidth]{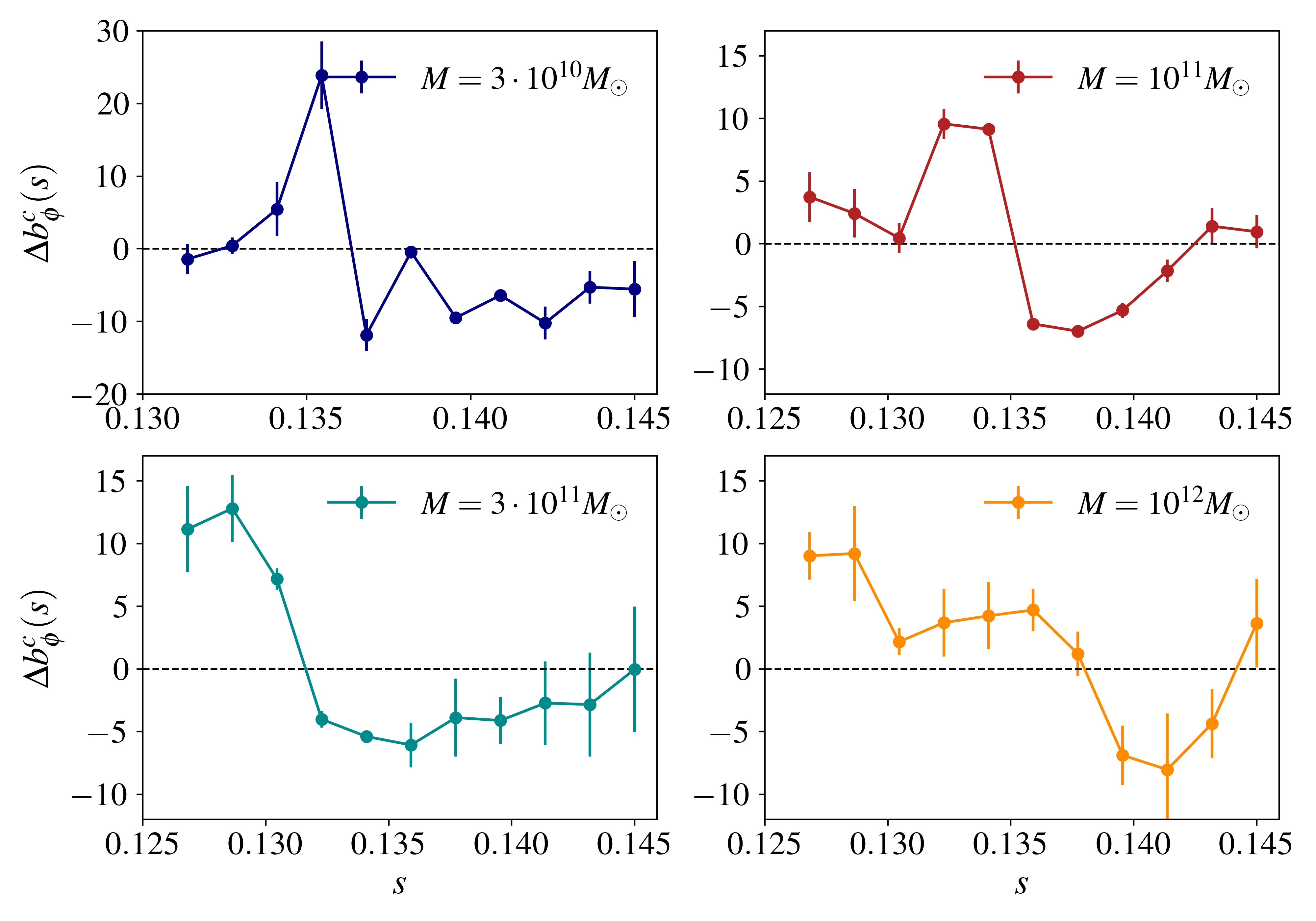}
    \includegraphics[width=0.48\textwidth]{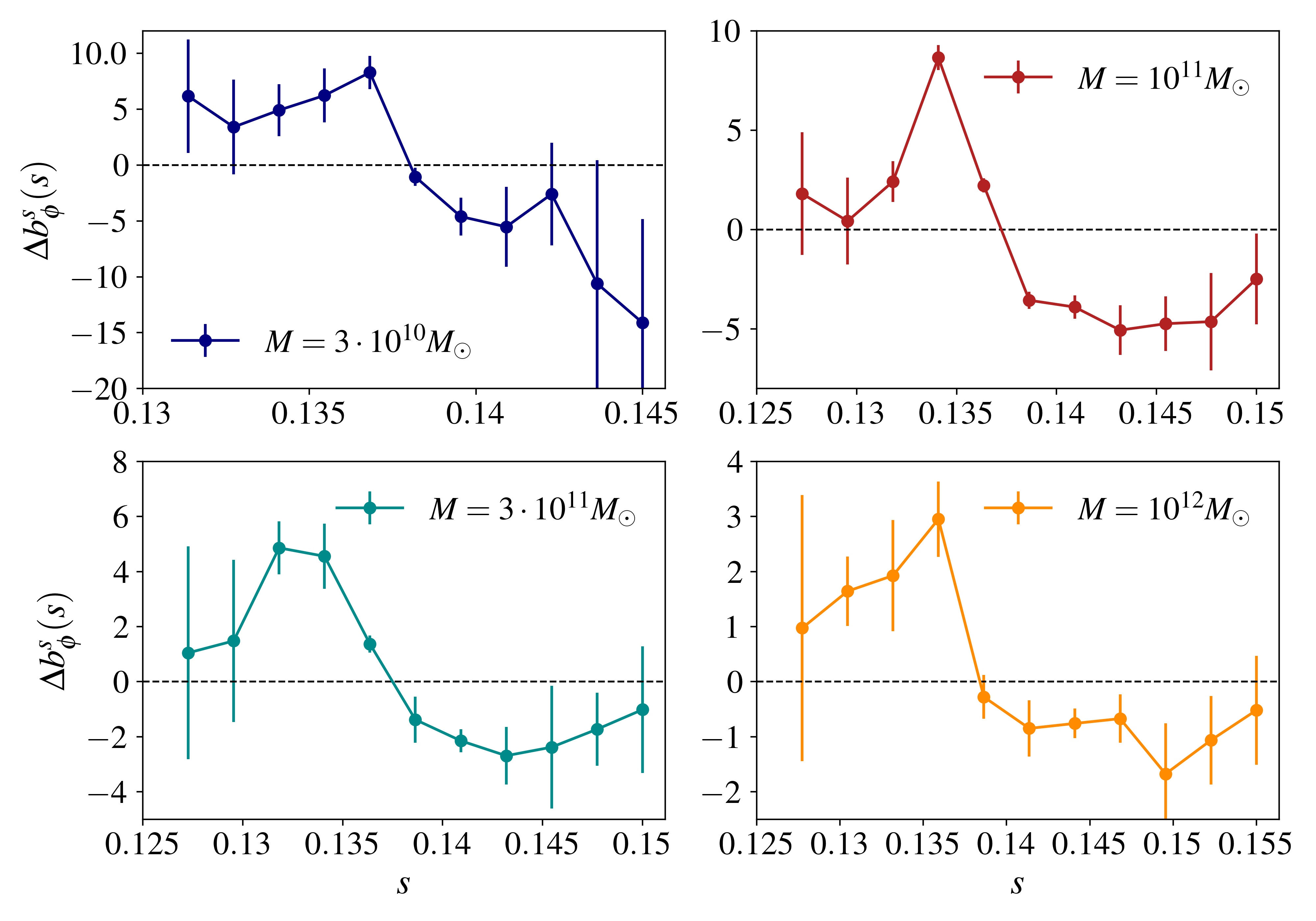}
    \caption{Same as Fig.~\ref{fig:bphiMs} but for galaxies selected according to their bulge-to-disk ratio $s$.}
    \label{fig:bphiShape}
\end{figure}

The correlation between a change in $\sigma_8$ and the halo formation time also appears to explain the dependence on the morphological parameter $s$ reported in ig.~\ref{fig:bphiShape}. 
Increasing $\sigma_8$ produces on average older galaxies, which are thought to have smaller disks~\cite{GalShape}, whence the observed positive values of $\Delta\bphi$ on the lower side of the $s$-range. Note that $s$ varies at most by 20\% at fixed $M_h$ for the halo masses considered here.

\section{Parameters for different galaxy formation model}
\label{app:gal}

\Gal provides a specific model for the different components that enter into galaxy formation, see~\cite{galacticus}. In this work we changed some of the parameters to study their effect on assembly bias. In tab.~\ref{tab:gal_par} we show the parameters for the different model. There, EH stands for efficiency heating, ESF for efficiency of stellar formation and VSF indicates the characteristic velocity for stellar feedback. The fiducial values for all the parameters used in this analysis can be found on the GitHub repository of \Gal\footnote{For the full list of fiducial parameters used see \href{https://github.com/galacticusorg/galacticus/wiki/Constraints:-Baryonic-Physics}{here}.}.

\begin{table*}
    \centering
    \begin{tabular}{|c|c|c|c|c|}
       Model  & EH (BH component) & ESF (disk/spheroid) & VSF (disk/spheroid) \\
       \hline
       Fiducial   & 0.0001576 & 0.2537~/~0.003064 & 49.96 / 41.53\\
       Low AGN  & 0.00007880 & 0.2537~/~0.003064 & 49.96 / 41.53 \\
       High AGN  & 0.0003152 & 0.2537~/~0.003064 & 49.96 / 41.53 \\
       Low SF  & 0.0001576 & 0.25367~/~0.003064 & 25.00 / 25.00 \\
       High SF  & 0.0001576 & 0.25367~/~0.003064 & 99.91 / 83.06 \\
       Low SFR  & 0.0001576 & 0.1268~/~0.001532 & 49.96 / 41.53 \\
       High SFR  & 0.0001576 & 0.5073~/~0.006128 & 49.96 / 41.53 \\
       \hline
    \end{tabular}
    \caption{Values for the parameters in the different galaxy formation models. These parameters are used to produce the results presented in section~\ref{sec:galmod}.}
    \label{tab:gal_par}
\end{table*}

\bibliographystyle{mnras}
\bibliography{biblio}

\label{lastpage}

\end{document}